%% file: main.tex
\documentclass{vldb_tech}

\def\BibTeX{{\rm B\kern-.05em{\sc i\kern-.025em b}\kern-.08em
    T\kern-.1667em\lower.7ex\hbox{E}\kern-.125emX}}

\usepackage{graphicx}
\usepackage{multicol}
\usepackage{balance}  
\usepackage{mathptmx}
\usepackage{booktabs} 

\usepackage{amssymb}
\usepackage{lscape}
\usepackage{array}
\usepackage{xcolor}
\usepackage{algorithmic}
\usepackage{subfigure}
\usepackage{amsmath}
\usepackage{mathrsfs}
\usepackage{float}
\usepackage{subfigure}
\usepackage{graphicx}
\usepackage[ruled,linesnumbered,lined,noend]{algorithm2e}
\usepackage{url}
\usepackage{textcomp}
\usepackage{enumitem}
\usepackage{mathtools}

\usepackage{tikz}
\usepackage{pgfplots}
\usepackage{pgfplotstable}
\usepackage{booktabs}
\usepackage{multirow}

\makeatletter
\newcommand{\removelatexerror}{\let\@latex@error\@gobble}
\makeatother

\pgfplotsset{
	every axis/.append style={
		legend columns=3,
		legend style={
			at={(0.5,1)},
			anchor=south,
			font=\small,
			draw=none,
			fill=none,
			row sep=0.01cm
		},
		height = 3.5cm,
		width = 5.3 cm,
		xlabel near ticks,
		ylabel near ticks,
		legend cell align = left,
		label style = {font=\small},
	},
	every tick label/.append style={
		font=\small
	},
}

\renewcommand{\textrightarrow}{$\rightarrow$}

\SetAlCapNameFnt{\small}
\SetAlCapFnt{\small}
\SetAlFnt{\small}
\DontPrintSemicolon
\SetKwInOut{KwData}{input}
\SetKwInOut{KwResult}{output}
\SetInd{5pt}{5pt}
\IncMargin{12pt}

\graphicspath{{figs/}}

\newcommand{\eat}[1]{}
\newcommand{\rbox}{\hfill $\Box$}

\newtheorem{definition}{Definition}[section]
\newtheorem{theorem}{Theorem}[section]
\newtheorem{lemma}[definition]{Lemma}

\newtheorem{example}[definition]{Example}

\newcommand{\paddingT}{\vskip -0.125in}
\newcommand{\paddingD}{\vskip -0.13in}

\newcommand\upA{\mathord{\uparrow}}
\newcommand\downA{\mathord{\downarrow}}

\newcommand{\colB}[1]{\textcolor{blue}{#1}}

\newcommand{\eatproofs}[1]{#1}
\newcommand{\eatTRstatement}[1]{}

\vldbTitle{ABC of Order Dependencies}
\vldbAuthors{Pei Li, Jaroslaw Szlichta, Michael B\"ohlen, Divesh Srivastava}
\vldbDOI{https://doi.org/10.14778/xxxxxxx.xxxxxxx}
\vldbVolume{13}
\vldbNumber{xxx}
\vldbYear{2020}

\begin{document}


\title{ABC of Order Dependencies}


\numberofauthors{4} 
\author{
\alignauthor
Pei Li\\
       \affaddr{Onedot A.G., Switzerland}\\
       \email{pei.li@onedot.com}
\alignauthor
Jaroslaw Szlichta\\
       \affaddr{Ontario Tech Univ, Canada}\\
       \email{jarek@ontariotechu.ca}
\alignauthor 
Michael B\"ohlen \\
       \affaddr{Univ of Z\"urich, Switzerland}\\
       \email{boehlen@ifi.uzh.ch}
\and  
\alignauthor Divesh Srivastava\\
       \affaddr{AT\&T Labs-Research, US}\\
       \email{divesh@research.att.com}
}


\maketitle

\begin{abstract}
 We enhance constrained-based data quality with approximate band conditional order dependencies (abcODs). Band ODs model the semantics of attributes that are monotonically related with small variations without there being an intrinsic violation of semantics. The class of abcODs generalizes band ODs to make them more relevant to real-world applications by relaxing them
 to hold approximately (abODs) with some exceptions and conditionally (bcODs) on subsets of the data. 
 We study the problem of automatic dependency discovery over a hierarchy of abcODs.
 
 First, we propose a more efficient algorithm to discover abODs than in recent prior work. The algorithm is based on a new optimization to compute  a longest monotonic band (longest subsequence of tuples that satisfy a band OD) through dynamic programming by decreasing the runtime from $O(n^2)$ to $O(n \log n)$ time. We then illustrate that while the discovery of bcODs is relatively straightforward, there exist codependencies between approximation and conditioning that make the problem of abcOD discovery challenging. The naive solution is prohibitively expensive as it considers all possible segmentations of tuples resulting in exponential time complexity. To reduce the search space, we devise a dynamic programming algorithm for abcOD discovery that determines the optimal solution in $O(n^3 \log n)$ complexity. To further optimize the performance, we adapt the algorithm to cheaply identify consecutive tuples that are guaranteed to belong to the same band--this improves the performance significantly in practice without losing optimality. While unidirectional abcODs are most common in practice, for generality we extend our algorithms with both ascending and descending orders to discover bidirectional abcODs. Finally, we perform a thorough experimental evaluation of our techniques over real-world and synthetic datasets.
 
\end{abstract}

\input{intro.tex}
\input{frame.tex}

\input{lmbs.tex}

\input{conditionalBand}

\input{discovery.tex}

\input{pieces.tex}
\input{bidirectional.tex}
\input{exp.tex}

\input{related.tex}

\balance

\bibliographystyle{abbrv}
\bibliography{main} 
\end{document}

%% file: intro.tex
\section{Introduction}
\label{sec:introduction}
\subsection{Motivation}\label{sec:mot}

Modern data-intensive applications critically rely on high quality
data to ensure that analyses are meaningful and do not fall prey to
the garbage in, garbage out (GIGO) syndrome.  In constraint-based data
quality, dependencies are used to formalize data quality requirements.
Previous work has focused mainly on functional dependencies
(FDs)~\cite{DBLP:journals/cj/HuhtalaKPT99}.  Several extensions to the
notion of an FD have been studied, including \emph{order dependencies}
(ODs)~\cite{SGG18,Langer:2016:EOD:2907337.2907581,DBLP:journals/pvldb/SzlichtaGGKS17},
which express rules involving order.



We introduce a novel data dependency \emph{approximate band
    conditional OD} (abcOD). \emph{Band ODs} express order
  relationships between attributes with small variations causing rules involving order including
  ODs~\cite{Langer:2016:EOD:2907337.2907581,DBLP:journals/pvldb/SzlichtaGGKS17,SGG18},
  sequential dependencies~\cite{Golab:2009:SD:1687627.1687693} and
  denial constraints~\cite{CIP13} to be violated without actual
  violation of application semantics. To match real world scenarios, we allow band ODs to hold either \emph{approximately} (abODs)~\cite{LSBS20} with some exceptions, \emph{conditionally} (bcODs) over subsets of the data, or both approximately and conditionally (abcODs) with a mix of
\emph{ascending} and \emph{descending} orders (bidirectional abcODs).

\eat{Being able to discover automatically BODs between ordered
  attributes helps to improve the data quality that is essential to
  many real-world applications.}\eat{ For example, having complete and
  accurate temporal information can significantly improve search
  engine optimization for e-commerce businesses, lead to better
  negotiated product prices for car purchasing and customer behavior
  prediction for airline reservations.} \eat{Therefore in this paper
  we study the problem of {\em discovering BODs and its application in
    repairing ordered attributes}.}


Table~\ref{tab:order} contains $22$ sample releases of the
\emph{Music} dataset (Reprise records) from
Discogs\footnote{\url{www.discogs.com}}\label{footnote:discogs} that are integrated from various sources. For
tracking purposes music companies assign a catalog number
($\sf{cat\#}$) to each release of a particular label.  {
  When lexicographically ordered by attribute ${\sf cat\#}$, the
  release date (encoded using attributes ${\sf year}$ and
  ${\sf month}$) is also approximately ordered over subsets of the
  data.
\begin{table}[t]\centering
 \vspace{-0.15in}
  \caption{\small{Reprise records.}
    }
  \label{tab:order}
  \scalebox{0.91}{
    \begin{tabular}{| l l l l r l  | }
      \hline
      {\sf id} & {\sf release}
      & {\sf country} & {\sf year} & {\sf month} 
       & {\sf cat\#} 
       \\
      \hline
      $t_1$ & Unplugged  & Canada & 1992 & Aug & CDW45024 
       \\
      $t_2$ & Mirror Ball  & Canada & \textbf{2012} & Jun & CDW45934 
        \\
      $t_3$ & Ether  & Canada & 1996 & Feb & CDW46012 
        \\
      $t_4$ & Insomniac & Canada & 1995 & Oct & CDW46046 
       \\
      $t_5$ & Summerteeth & Canada & 1999 & Mar & CDW47282 
       \\
      $t_6$ & Sonic Jihad  & Canada & 2000 & Jul & CDW47383 
       \\
      $t_{7}$ & Title of...  & Canada & 1999& Jul  & CDW47388 
       \\
      $t_{8}$ & Reptile  & Canada & 2001 & Mar & CDW47966 
       \\
      $t_{9}$ & Always... & Canada & 2002 & Feb & CDW48016 
       \\
      \hline
      $t_{10}$ & Take A Picture  & US & 2000 & Nov & 9 16889-4 
      \\
      $t_{11}$ & One Week  & US & 1998 & Sep & 9 17174-2 
       \\
      $t_{12}$ & Only If...  & US & 1997 & Nov & 9 17266-2 
       \\
      $t_{13}$ & Never...  & US & 1996 & Nov & 9 17503-2 
       \\
      $t_{14}$ & We Run ...  & US & 1994 & Dec & 9 18069-2 
       \\
      \hline
      $t_{15}$ & The Jimi...  & US & 1982 & Aug & 9 22306-1 
       \\
      $t_{16}$ & Never...  & US & 1987 & Jan &  9 25619-1 
        \\  
       $t_{17}$ & Vonda Shepard & US & 1989 & Mar  & 9 25718-2 
        \\   
      $t_{18}$ & Ancient Heart  & US & \textbf{Null} & Jul &  9 25839-2 
       \\
      $t_{19}$ & Twenty 1  & US & 1991 & May & 9 26391-2  
        \\
      $t_{20}$ & Stress  & US & 1990  & Apr &  9 26519-1 
       \\
            $t_{21}$ & Play & US & 1991 & Mar &  9 26644-2 
             \\
            $t_{22}$ & Handels... & US & 1992 & Apr &  9 26980-2 
             \\
      \hline
    \end{tabular}
    }
\paddingD
\end{table}
Release dates 
are approximately ordered within subsets of the tuples called
\emph{series}, i.e., $\{t_1$--$t_{9}\}, \{t_{10}$--$t_{14}\}$ and
$\{t_{15}$--$t_{22}\}$ modulo ascending and descending orders.

Note that tuple $t_3$ has a smaller ${\sf cat\#}$ than $t_4$
(${\sf CDW46012}$ $<$ ${\sf CDW46046}$), however, is released a few months
later than tuple $t_4$ ($1996\mbox{/Feb} > 1995\mbox{/Oct}$; for
  month the sort order is according to the calendar ordering). This
is common in the music industry as ${\sf cat\#}$ is often assigned to
a record before it is released at the production stage.
Thus, tuples with delayed release dates will slightly violate an OD
between ${\sf cat\#}$ and (${\sf year}$, ${\sf month}$). A permissible
range to accommodate these small variations is called a \emph{band}.

Attribute ${\sf year}$ has also a missing value (tuple $t_{18}$) and
an erroneous value (tuple $t_2$) that severely break the OD between
${\sf cat\#}$ and (${\sf year}$, ${\sf month}$), as the value of ${\sf year}$ for
tuple $t_2$ is $2012$ and for tuple $t_3$ is $1996$
despite the ascending trend within the series. We verified that the
correct value of ${\sf year}$ for tuple $t_2$ should be $1995$.
(Table~\ref{tab:motivation} shows statistics of
violations.) 

}

\begin{table}[t]\centering
  \caption{\small{Statistics of top-5 music labels of {\em Discogs}.}}
  \label{tab:motivation}
  \scalebox{0.81 }{
  \small
    \begin{tabular}{|c|c|c|c|}
      \hline
      {\sf \textbf{label}} & {\sf \textbf{\# total releases}}
      & {\sf \textbf{\# missing years}}
      & {\sf \textbf{\# incorrect years}}
      \\ \hline
           Capitol Records                & 28935 & 3392 & 896 \\ \hline
            Reprise Records               & 9830 & 688& 304\\ \hline
            Ninja Tune               & 2055 & 10 & 33 \\ \hline
            V2 Records               & 1551  & 13 & 15 \\ \hline
             BGO Records              & 588 & 47 & 13 \\ \hline
    \end{tabular}
  }
\paddingD
\end{table}

\begin{figure}[t]\centering
  \includegraphics[scale=0.3]{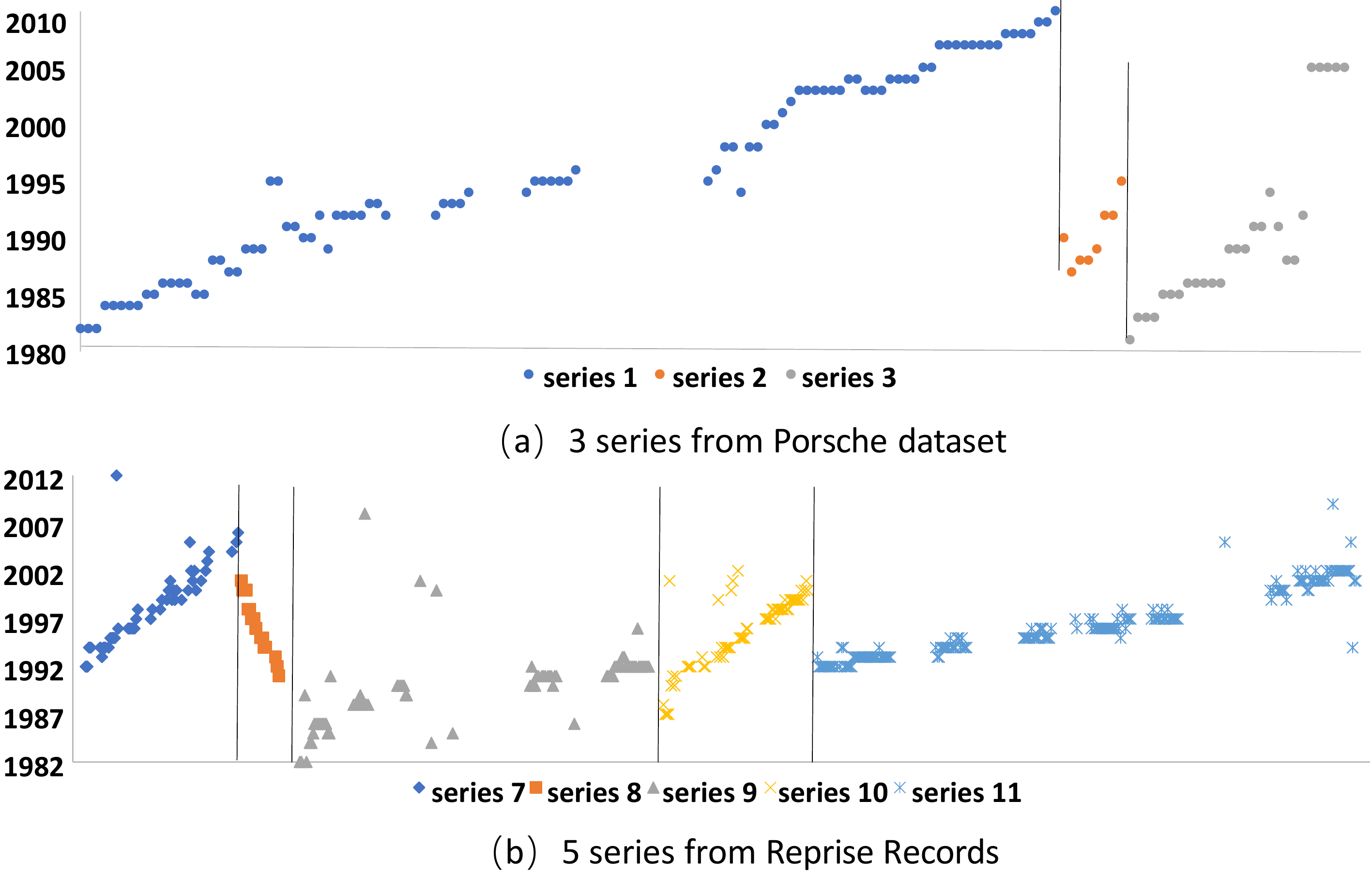}
   \paddingT
  \caption{\small{Real-world series in \emph{Car} and \emph{Music}
      datasets.}}
    \paddingD
  \label{fig:real-world}
\end{figure}

{

  As another example, since vehicle identification numbers (${\sf VINs}$) for
  cars are assigned sequentially and independently by different manufacturing plants, attributes ${\sf VIN}$ and
  ${\sf year}$ in car datasets are conditionally ordered over subsets
  of data. There are small variations to the OD between these
  attributes as ${\sf VIN}$s are assigned to a car before it is
  manufactured and ${\sf year}$ denotes the time of the completion of
  the product. There are also actual errors to this band OD (as illustrated
  in Figure~\ref{fig:real-world}a), due to data quality issues.
  Fig.~\ref{fig:real-world} plots a small sample of the real-world
  \emph{Car} dataset%
  \footnote{\url{www.classicdriver.com}} and \emph{Music} dataset
  series (separated by vertical lines). Series are identified by
  ${\sf VIN}$ and ${\sf cat\#}$, respectively.
}


Data dependencies to identify data quality errors can be obtained
manually through a consultation with domain experts, however, this is known
to be an expensive, time consuming, and error-prone
process~\cite{Golab:2009:SD:1687627.1687693,DBLP:journals/cj/HuhtalaKPT99,DBLP:journals/pvldb/SzlichtaGGKS17}. Thus,
automatic approaches to discover data dependencies to identify data
quality issues are needed.
The key technical problem that we study is how to automatically and efficiently discover data dependencies over a \emph{hierarchy} of abcODs classes. While \emph{band} ODs to permit small variations or \emph{approximate} band ODs (abODs) to accommodate errors might be sufficient in some applications, there is also
a need to study band \emph{conditional} ODs (bcODs) and \emph{approximate} band \emph{conditional} ODs (abcODs) that hold over subsets of the data. Unidirectional abcODs are most common in practice (as we verified experimentally in Sec.~\ref{sec:experimental_evaluation}), however, in some applications to understand the data one needs the additional semantics of \emph{ascending} and \emph{descending} orders that the \emph{bidirectional} abcOD captures. Automatically discovered abcODs can be manually validated by domain experts, which is a much easier task than manual specification, and identified errors corrected subsequently. 

\subsection{Contributions}
\label{sub:contributions}

There are two variants of data dependency discovery algorithms. The
first one is a global approach to automatically find all dependencies
that hold in the
data~\cite{DBLP:journals/cj/HuhtalaKPT99,SGG18,Langer:2016:EOD:2907337.2907581,DBLP:journals/pvldb/SzlichtaGGKS17}. The
second one is a relativistic approach to find subsets of the data
obeying the expected
semantics~\cite{Golab:2009:SD:1687627.1687693,Golab:2008:GNT:1453856.1453900}, which is
laborious to do manually. We apply a \emph{hybrid} approach to the
discovery of abcODs that combines these two approaches.

To automatically identify candidates for embedded band ODs without
human intervention, we use a cheaper global approach that finds all
traditional ODs within an approximation
ratio~\cite{SGG18,DBLP:journals/pvldb/SzlichtaGGKS17}.  The approach
in \cite{SGG18,DBLP:journals/pvldb/SzlichtaGGKS17} is limited to
identifying ODs that (i) do not permit small variations within a band,
thus, we deliberately set the approximation ratio higher, and (ii)
hold over the entire dataset rather than subsets of the data, thus, we
separate the data into segments by using a \emph{divide-and-conquer}
approach. 
We use the identified traditional ODs, ranked by the measure of
interestingness~\cite{SGG18,DBLP:journals/pvldb/SzlichtaGGKS17}, as
candidate embedded band ODs to solve the problem of discovering
abcODs.

We define the problem of abcOD discovery as an \emph{optimization problem} desiring parsimonious number of segments that identify using conditions large fractions of the data (\emph{gain}) each of which satisfies the embedded band OD with few violations (\emph{cost}). 

We make the following contributions in this paper.

\begin{enumerate}[nolistsep,leftmargin=*]
\item \emph{Hierarchy of abcOD classes.} 
  \begin{enumerate}[nolistsep,leftmargin=*]
    \item We introduce a novel \emph{approximate band conditional OD} (abcOD) integrity constraint. Band ODs are based on small variations causing traditional ODs to be violated without an actual violation of application semantics. abcODs generalize band ODs to make them more applicable to real-world data by relaxing their requirements to hold \emph{approximately} (abODs)~\cite{LSBS20} with some exceptions and \emph{conditionally} (bcODs) on subsets of the data. abcODs do not consider a mix of \emph{ascending} and \emph{descending} orders, as \emph{bidirectional} abcODs do.
    \item We develop methods to automatically compute the \emph{band-width} to allow for small variations and \emph{candidate dependencies} for the hierarchy of (bidirectional) abcODs to decrease the human burden of specifying them manually.
  \end{enumerate}
\item \emph{Discovery over hierarchy of abcODs.}
    \begin{enumerate}[nolistsep,leftmargin=*]
    \item We improve on the recent work on the abODs discovery~\cite{LSBS20} based on the notion of a \emph{longest monotonic band} (LMB) to identity longest subsequences of tuples that satisfy a band OD. We provide a new optimization to the dynamic programming algorithm to improve the LMB computation from $O(n^2)$ to $O(n \log n)$ time, where $n$ is the number of tuples.
    \item We show that while the discovery of bcODs is relatively straightforward, there are codependencies between approximation and conditioning that introduce new challenges to the abcOD discovery problem. The naive solution to consider all possible segmentations of tuples is prohibitively expensive, as it leads to exponential time complexity. We devise a \emph{dynamic programming} algorithm based on LMBs that finds the optimal solution in $O(n^3 \log n)$ time. To further decrease the search space, we optimize the algorithm without losing optimality based on \emph{pieces}, which are contiguous subsequences of tuples that satisfy band monotonicity. The optimized pieces-based algorithm is $O(m^2 n \log n)$ time, where $m$ is the number of pieces. Since in practice pieces are large, hence, the number of pieces is much smaller than the number of tuples, the pieces-based algorithm is orders-of-magnitude faster. 
    \item We extend our algorithms to account for ascending and descending orders. Based on this, we devise algorithms to discover \emph{bidirectional} abcODs. 
    \end{enumerate}
    
\item \emph{Experiments.}
We experimentally demonstrate the effectiveness and scalability of our solution, and compare our techniques with baseline methods on real-world and synthetic datasets.
\end{enumerate}


We provide basic definitions in Section~\ref{sec:overview}. 
In Sections \ref{sec:longest_increasing_band}--\ref{sec:bidDiscovery}, we study algorithms to discover abODs, abcODs and bidirectional abcODs, respectively. 
We discuss experimental results in Section~\ref{sec:experimental_evaluation} and related work in Section~\ref{sec:related_work}. We conclude in Section~\ref{sec:conclusion}. 

%% file: frame.tex
\section{Background} \label{sec:overview}

{
We use the following notational conventions.
\begin{itemize}
\item \textbf{Relations.}  $R$ denotes a relation schema and $r$
  denotes a specific table instance. Italic letters from the beginning
  of the alphabet $A, B$ and $C$ denote single attributes. Also, $s$
  and $t$ denote tuples in $r$ and $s.A$ denotes the value of an
  attribute $A$ in a tuple $s$.
  $\sc{dom}\mbox{(}A\mbox{)}$ denotes the domain of an attribute $A$.
\item \textbf{Lists.}  Bold letters from the end of the alphabet
  \textbf{X, Y} and \textbf{Z} denote lists of attributes.
  $\mbox{[}A, B, C\mbox{]}$ denotes an explicit list of attributes.
  $\sc{dom}\mbox{(}\textbf{X}\mbox{)} = \sc{dom}\mbox{(}A\mbox{)}
  \cdot \sc{dom}\mbox{(}B\mbox{)} \cdot \sc{dom}\mbox{(}C\mbox{)}$
  denotes the domain of $\textbf{X}$, where
  $\textbf{X} = \mbox{[}A, B, C\mbox{]}$.
            $s.\textbf{X}$ denotes the value of the list of attributes $\textbf{X}$ in the tuple $s$.
          \end{itemize}
}

{

  Let
  $d$: $\sc{dom}\mbox{(}\textbf{X}\mbox{)} \cdot
  \sc{dom}\mbox{(}\textbf{X}\mbox{)} \rightarrow \mathbb{R}$ be a
  \emph{distance function} defined on the domain of
  $\textbf{X}$. Distance function $d$ satisfies the following
  properties: \emph{anti-symmetry, triangle inequality} and
  \emph{identity of indiscernibles}. We consider $d\mbox{(}x_1, x_2\mbox{)} = || x_2 || - || x_1 ||$, where $|| x ||$ denotes the norm of the value list $x$.
\eat{
\begin{itemize}
\item anti-symmetric (???): $d(x_1, x_2) = - d(x_2, x_1)$;
\item triangle inequality: $d(x, z) \leq d(x, y) + d(y, z)$;
\item identity of indiscernible: $d(x, y) = 0 \Leftrightarrow x = y$. 
\end{itemize}
}
%
%
We model an \emph{order specification} as a directive to sort a dataset in \emph{ascending} or \emph{descending} order.

\begin{definition}[Order Specification]
  An order specification is a marked list of  attributes, denoted as
  $\overline{\textbf{Y}}$. There are two ordering directions:
  ${\sf asc}$ and ${\sf desc}$, indicating ascending and descending
  ordering, respectively. As shorthand, $\textbf{Y} \upA$ indicates
  $\textbf{Y}$ ${\sf asc}$ and $\textbf{Y} \downA$ indicates
  $\textbf{Y}$ ${\sf desc}$. \rbox
\end{definition}

\begin{definition}[Order Operator]
  Let $\overline{\textbf{Y}}$ be a marked list of attributes in a
  relation $r$ and let $\Delta$ be a constant value.
  For two tuples $t, s \in r$,
  $t \preceq_{\Delta, \overline{\textbf{Y}}} s$ if
  \begin{itemize}[nolistsep]
  \item $\overline{\textbf{Y}} =$ $\textbf{Y} \upA$ and
    $d(t.\textbf{Y}, s.\textbf{Y}) \geq -\Delta$; or
  \item $\overline{\textbf{Y}} =$ $\textbf{Y} \downA$ and
    $d(s.\textbf{Y}, t.\textbf{Y}) \leq \Delta$.
  \end{itemize}
  Let $t \prec_{\Delta, \overline{\textbf{Y}}} s$ if $t \preceq_{\Delta, \overline{\textbf{Y}}} s$ but $s \not \preceq_{\Delta, \overline{\textbf{Y}}} t$ and 
  let $t \preceq_{\overline{\textbf{Y}}} s$ be the operator
  $t \preceq_{\Delta, \overline{\textbf{Y}}} s$, where $\Delta =
  0$. \rbox
\end{definition}

\begin{definition}[Bidirectional Band OD]
  Given a band-width $\Delta$,
  a list of attributes $\textbf{X}$, a marked list of attributes
  $\overline{\textbf{Y}}$ over a relation schema $R$, a bidirectional \emph{band
    order dependency} (bidirectional band OD) denoted by
  $\textbf{X} \mapsto_{\Delta} \overline{\textbf{Y}}$ holds over a
  table $r$, if $t \preceq_{\textbf{X}\upA{}} s$ implies
  $t \preceq_{\Delta, \overline{\textbf{Y}}} s$ for every tuple pair
  $t, s \in r$.
  \rbox
  \label{def:bandOD}
\end{definition}

\begin{example}\label{example:bandODSer}
  A bidirectional band OD $\sf{cat\#} \mapsto_{\Delta=1} \sf{year}\upA$ holds over
  tuples $\{t_1, t_3$--$t_9\}$ in Table~\ref{tab:order} with a
  band-width of one year.
%
%
A bidirectional band OD
$\sf{cat\#} \mapsto_{\Delta=12} [\sf{year},
\sf{month}]\upA$ holds over tuples $\{t_1, t_3- t_9\}$ in
Table~\ref{tab:order} with a band-width of 12 months.
\rbox
\end{example}

Bidirectional band ODs specify that when tuples are ordered increasingly on the left-hand-side ($\sf{cat\#}$ in
  Example~\ref{example:bandODSer}), their right-hand-side ($\sf{year}$ in
  Example~\ref{example:bandODSer}) must be ordered non-decreasingly
(e.g., wrt series $S_1$ and $S_3$ in
  Example~\ref{example:bandODSeries} discussed next) or non-increasingly
(e.g., wrt the series $S_2$ in
  Example~\ref{example:bandODSeries}) within the specified band-width
(e.g., $\Delta = 1$ in Example~\ref{example:bandODSer}).

Since both ascending and descending trends are allowed, the consequent
of the dependency is a list of marked attributes. We describe how to
automatically compute band-width in
Sec.~\ref{sub:estimating_parameters}.  Note that traditional bidirectional 
ODs~\cite{Langer:2016:EOD:2907337.2907581,SGG18,DBLP:journals/pvldb/SzlichtaGGKS17, Szlichta:2012:FOD:2350229.2350241,Szlichta:2013:ECO:2556549.2556568}
are a special case of bidirectional band ODs, where $\Delta = 0$. We support
  other data types than numerical columns including categorical
  columns. For instance, months can be represented as strings as in
  the example in Section~\ref{sec:mot} over
  Table~\ref{tab:order}. Whenever the distance function can be
  preserved, the values of the columns are replaced with integers: 1,
  ..., n, in a way that keeps the same ordering, i.e., higher values
  are replaced by larger integers. Computation over integers is more
  time and space efficient.

We consider a subclass of bidirectional band ODs, called  simply \emph{band ODs} for which bidirectionality is removed. We verified experimentally that unidirectional band ODs are most common in real-life applications in Section ~\ref{sec:experimental_evaluation}.

\begin{definition}[Band OD]
A  bidirectional band OD  is  call\-ed a band order  dependency (band OD)  when a list of attributes within it is all marked as ${\sf asc}$ or all as ${\sf desc}$.
\end{definition}


In real-world applications, (bidirectional) band ODs often hold \emph{approximately} with
some exceptions to accommodate errors and \emph{conditionally} over
subsets of the data (\emph{series}). We call these dependencies \emph{(bidirectional) approximate conditional band ODs} ((bidirectional) abcODs).
\begin{example}\label{example:bandODSeries}
  As shown in Figure~\ref{fig:series_example}, there are three series in Table~\ref{tab:order}:
  $S_1=\{t_1$---$t_{9}\}$ over bidirectional abcOD ${\sf cat\#}$ $\mapsto_{\Delta = 1}$ ${\sf year} \upA$, $S_2=\{t_{10}$--$t_{14}\}$  over bidirectional abcOD ${\sf cat\#}$ $\mapsto_{\Delta = 1}$ ${\sf year} \downA$ and
  $S_3=\{t_{15}$--$t_{22}\}$ over bidirectional abcOD ${\sf cat\#}$ $\mapsto_{\Delta = 1}$ ${\sf year} \upA$.  There is a tuple with an erroneous ${\sf
    year}$ ($t_2$; correct ${\sf year}$ is 1995), and a tuple with a
  missing ${\sf year}$ ($t_{18}$; correct ${\sf year}$ is 1988). \rbox
\end{example}

We desire parsimonious segments that identify large subsets of data that satisfy a (bidirectional) band OD with few violations. We formally define the discovery problem over a hierarchy of bidirectional abcODs
as an optimization problem in Sections~\ref{sec:longest_increasing_band}--\ref{sec:bidDiscovery}.

}

%% file: lmbs.tex
\section{Discovery of \MakeLowercase{ab}OD\MakeLowercase{s}}
\label{sec:longest_increasing_band}

In order to reduce the search space of the data dependencies discovery problem, we use the notion of a \emph{longest monotonic band} (LMB) to identify the longest subsequences of tuples that satisfy a band OD. We formally define LMBs in Sec.~\ref{sub:definition}, then present key properties of LMBs, which lead to efficient calculation of LMBs in Sec.~\ref{sub:computing_sf_lib_}. We use LMBs in Sec.~\ref{sub:estimating_parameters} to automatically compute the optimal band-width. (Thus, Sec.~\ref{sub:estimating_parameters} is presented after Sections~\ref{sub:definition}--~\ref{sub:computing_sf_lib_}). The computation of LMBs and band-widths are used to discover abODs in Section~\ref{sub:abOD_discovery} as well as to discover bcODs and (bidirectional) abcODs in Sections~\ref{sec:series}--\ref{sec:bidDiscovery}. Since \emph{unidirectional} abcODs are most common in practice as shown in Table~\ref{tab:stis}, we focus on them in Sections~\ref{sec:longest_increasing_band}--\ref{sec:series}. Without loss of generality, we use ascending order as default, i.e., a marked list of attributes $\overline{\textbf{Y}}$ is cast to $\textbf{Y} \upA$ when unidirectionality is expected. (We present how to extend our algorithms to cover bidirectionality in Section~\ref{sec:bidDiscovery}.)


  
\subsection{Defining LMB}
\label{sub:definition}

\begin{figure}[t] \centering
  \includegraphics[scale=0.55]{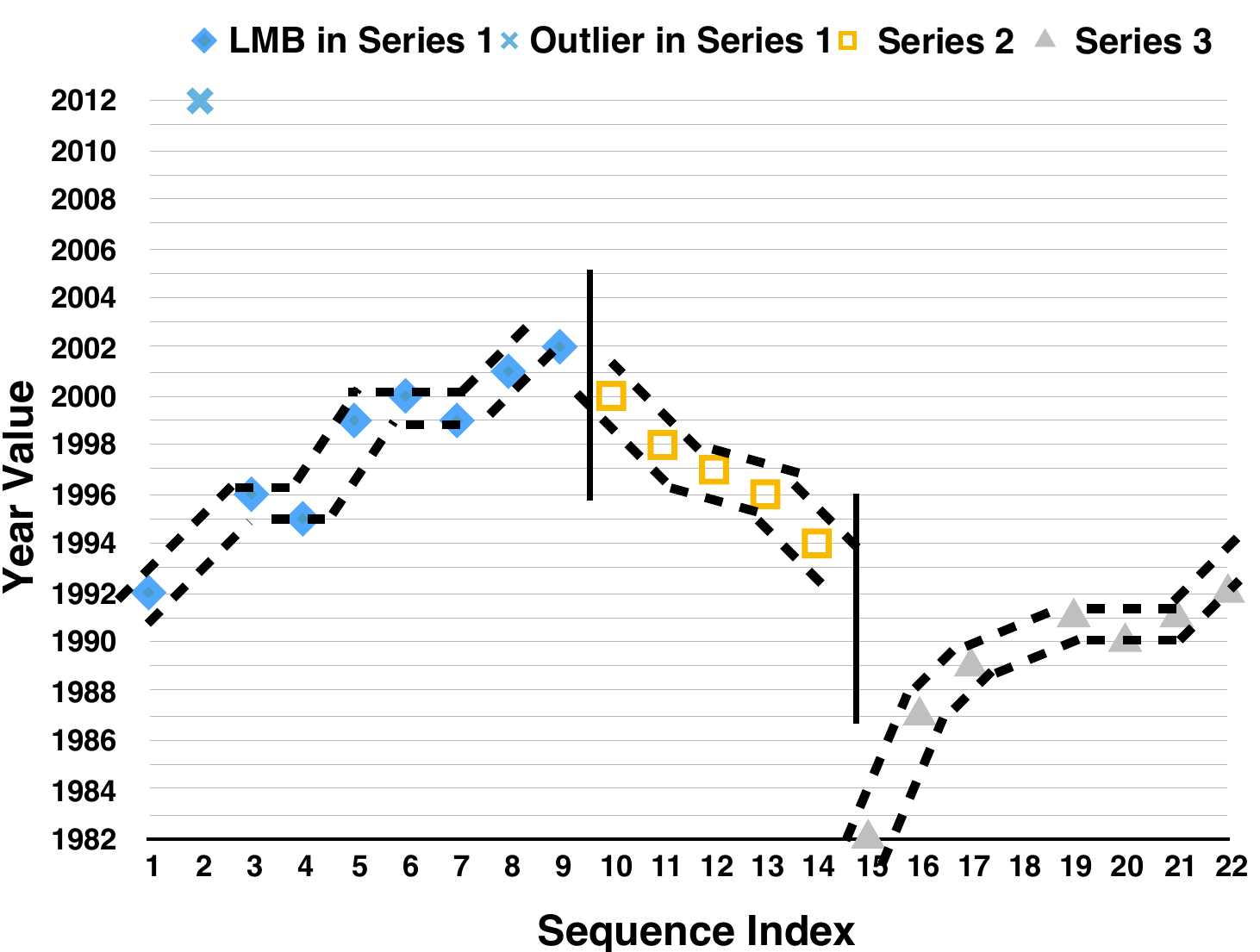}
\paddingT
  \caption{\small{Determining abcODs in Table~\ref{tab:order}, $\Delta$=1.
     \label{fig:series_example}}}
     \paddingD
\end{figure}

In contrast to previous work on identifying a longest monotonic subsequence~\cite{Albert2004405,Chen2007223,CROCHEMORE:2008:CLI:2227536.2227543,Golab:2009:SD:1687627.1687693,DBLP:journals/jco/Liben-NowellVZ06} 
the definition of a longest monotonic band~\cite{LSBS20} allows for \emph{slight variations}. Recall, that when we consider a band OD $\sf{cat\#} \mapsto_{\Delta=1} \sf{year} \upA$ and a series $S_1=\{t_1-t_{9}\}$ in Fig.~\ref{fig:series_example} over Table~\ref{tab:order}, tuples $t_4$ and $t_7$ are considered to be correct (as they are within the band-width) and only tuple $t_2$ is incorrect. LMBs are defined with respect to a band OD $\textbf{X} \mapsto_{\Delta} {\overline{\textbf{Y}}}$. In the remaining, $T=\{t_1, \cdots, t_n\}$ denotes a sequence of tuples ordered lexicographically by $\textbf{X}$ in ascending order.


\begin{definition}[Longest Monotonic Band]\label{def:2}
  Given a\\ se\-quence of tuples $T=\{t_1, t_2, \cdots, t_n\}$, a marked list of attributes $\overline{\textbf{Y}}$ and band-width $\Delta$, a
  \emph{monotonic band} (MB) is a subsequence of tuples $M=$
  $\{t_i, \cdots, t_j\}$ over $T$, such that
  $\forall_{k_1, k_2 \in \{i, \cdots, j \}, k_1 < k_2 }$
  $t_{k_1} \preceq_{\Delta, \overline{\textbf{Y}}} t_{k_2}$. The longest subsequence $M$ satisfying this
  condition is called a longest monotonic band (LMB). \rbox
\label{def:lmbSeq}
\end{definition}

\begin{example}\label{ex:series}
  Consider a band OD ${\sf cat\#}$ $\mapsto_{\Delta = 1}$ ${\sf year} \upA$ over  Table~\ref{tab:order}. In 
  $T$ = $\{ t_{1}- t_{9} \}$ ordered by ${\sf cat\#}$, 
there is a LMB $\{t_1, t_{3}- t_{9} \}$ with the values of {\sf year}
  $\{'92, '96, '95, '99, '00, '99,$ $'01, '02\}$.
  Note that a LMB can have local decreases within the band-width 
  \eat{Based on Definition~\ref{def:lmbSeq}, since 
    $t_{10} \preceq_{\Delta=1, \sf{year}\upA{}} t_{11}$,
    $t_{11} \preceq_{\Delta=1,\sf{year}\upA{}} t_{13}$ and 
    $t_{11} \preceq_{\Delta=1, \sf{year}\upA{}} t_{14}$ 
do not hold, tuples $t_{10}$, $t_{13}$ and $t_{14}$ are not part of the LIB with tuples $t_{11}$ and $t_{12}$. However,  
    $t_{10} \preceq_{\Delta=1,\sf{year}\downA{}} t_{11}$,
    $t_{10} \preceq_{\Delta=1,\sf{year}\downA{}} t_{12}$,
    $t_{10} \preceq_{\Delta=1,\sf{year}\downA{}} t_{13}$,
    $t_{10} \preceq_{\Delta=1,\sf{year}\downA{}} t_{14}$,
    $t_{11} \preceq_{\Delta=1,\sf{year}\downA{}} t_{12}$, 
    $t_{11} \preceq_{\Delta=1,\sf{year}\downA{}} t_{13}$,
    $t_{11} \preceq_{\Delta=1,\sf{year}\downA{}} t_{14}$,
    $t_{12} \preceq_{\Delta=1,\sf{year}\downA{}} t_{13}$,
    $t_{12} \preceq_{\Delta=1,\sf{year}\downA{}} t_{14}$ and
    $t_{13} \preceq_{\Delta=1,\sf{year}\downA{}} t_{14}$. 
Thus, tuples $t_{10}$--$t_{14}$ form a LDB.}
and is also not necessarily a contiguous subsequence of tuples. \eat{For example, in Fig.~\ref{fig:series_example}
  a LIB over series with tuples $t_{1}$--$t_{9}$ includes tuples $t_{1}$ $\cup$  $t_{3}$--$t_{9}$. }
  \rbox
\end{example}

\eat{
\begin{definition}[Maximal Tuple]
  Given a sequence of tuples $T=\{t_1, \cdots, t_n\}$ and a list of
  attributes $\textbf{Y}$, a tuple $t_i \in T$ is a \emph{maximal
    tuple}, denoted as
    $\max_{\textbf{Y}}$($t_1, \cdots, t_n$), if 
  $\forall_{j \in \{1, \cdots, n \} }$
  $d\mbox{(} t_{j}.\textbf{Y}, t_{i}.\textbf{Y} \mbox{)}$ $\geq$ $0$. \rbox
\end{definition}

\begin{example}
Given $T = \{ t_{3}$--$t_{7}\}$ over Table~\ref{tab:order}, $t_{6}$ is a maximal tuple wrt ${\sf year}$. \rbox
\end{example}
}

\begin{definition}[Max \& Min Tuples]
  Given a sequence $T=\{t_1, \cdots, t_n\}$ and list of
  attributes $\textbf{Y}$, tuple $t_i \in T$ is a maximal
    tuple, denoted as
    $\max_{\textbf{Y}}$($t_1, \cdots, t_n$), if
  $\forall_{j \in \{1, \cdots, n \} }$
  $t_j \preceq_{\textbf{Y} \upA} t_i$
and a minimal tuple 
denoted as $\min_{\textbf{Y}}$($t_1, \cdots, t_n$),
if $\forall_{j \in \{1, \cdots, n \} }$ 
$t_i \preceq_{\textbf{Y} \upA} t_j$
\rbox
\end{definition}

\begin{example}
Given $T = \{ t_{3}$--$t_{7}\}$ over Table~\ref{tab:order}, $t_{4}$ is a minimal tuple and $t_{6}$ is a maximal tuple over ${\sf year}$. \rbox
\end{example}
	
\subsection{Computing LMB} 
\label{sub:computing_sf_lib_}

We proposed in prior work~\cite{LSBS20}, an algorithm to compute a LMB in $O(n^2)$ time by reducing the problem to finding LMBs in subsequences. Let $T[i]$ denote the prefix of a sequence $T$ of length $i$, i.e., $T[i] = \{t_i, t_2, \cdots, t_i\}$, where $i \in [1, n]$. Among all MBs that end at $t_i$ in $T[i]$, the longest one is kept including its maximal tuple. The solution is based on the optimal substructure property that in order to find a LMB in sequence $T[i+1]$, given those of $T[1]$ till $T[i]$, tuple $t_{i+1}$ is verified whether it can extend the length of any existing LMBs in $T[i]$ based on their maximal tuples. The MB with the longest length is kept.
Once LMBs in subsequences are enumerated, the longest one is chosen as a LMB.

\begin{example}
Assume $T=\{t_1$--$t_4\}$ over Table~\ref{tab:order}, $\textbf{Y} = [\sf{year}]$ and $\Delta =1$. Initially, first tuple is a LMB in $T[1]$ of length one with a maximal tuple $t_{1}$. Since $t_1 \preceq_{\Delta, \sf{year} \upA} t_2$, $t_2$ can extend the LMB in $T[1]$ of length two in $T[2]$ with a maximal tuple $t_2$. Similarly, tuple $t_3$ can extend the LMB in $T[1]$, however not the one in $T[2]$ as $t_3 \preceq_{\Delta, \sf{year} \upA} t_2$, hence, a LMB of length two in $T[3]$ with the maximal tuple $t_3$ is obtained. Finally, since tuple $t_4$ can extend LMBs in $T[1]$ and $T[3]$, but the one based on $T[3]$ is the longest, a LMB of length three in $T$ is obtained with a maximal tuple $t_3$. 
\end{example}


In this work, we propose an efficient algorithm to compute a LMB that improves the complexity to $O(n\log n)$ time. The new optimization to reduce the search space is based on maintaining only one monotonic band of each possible length with a minimal tuple among maximal tuples called a {\em best tuple}. This is based on the observation that tuple $t_{i+1}$ can extend at least one of the previous MBs of predefined length $k$ in all subsequences in $T[i]$ with best tuples denoted as $s_{k,i}$, if  $s_{k,i} \preceq_{\Delta, \textbf{Y} \upA} t_{i+1}$.


\begin{definition}[Best tuple]
Given a sequence of tuples $T$, band-width $\Delta$ and list of attributes $\textbf{Y}$, for each $k, i \in \{ 1, \cdots n \}$, $s_{k, i}$ is a best tuple of MBs of length $k$ in $T[i]$, if $s_{k, i}$ is a minimal tuple among maximal tuples of all {\sf MB}s with length $k$ in $T[i]$.	
  \rbox \label{def:besttup}
\end{definition}

\begin{example}\label{example:best_tuple}
%
Consider a sequence of tuples $T =\{t_1= '92$, $t_2$ $=$ $'12$, $t_3 = '95\}$ and band-width $\Delta =1$. There are three
  MBs of length one: $t_1, t_2$, and $t_3$ in $T$, among
  which $t_1$ is a best tuple. Similarly, there are
  two MBs of length two: $\{t_1, t_2\}$ with a maximal tuple $t_2$ 
  and
  $\{t_1,t_3\}$ with a maximal tuple $t_3$. 
  Thus, tuple $t_3$ is a best tuple of MBs of length two in $T$. \rbox
\end{example}

The following theorem 
is helpful to decide if tuple $t_{i+1}$ extends previous MBs of length $k$ in $T[i]$ to MB with a best tuple 
of lenght $k+1$ in $T[i+1]$ or a MB with a best tuple of length $k+1$ in $T[i]$ remains the MB with best tuple of length $k+1$ in $T[i+1]$. 
%
\eat{Without losing generality, we focus on {\sf LIB}s in the rest of this section.} 
\eatTRstatement{
Due to the space limit, the proofs of all theorems and lemmas can be found in the technical report~\cite{LBS19}.
}

\begin{theorem}\label{them:best}
Given a band-width $\Delta$, sequence of tuples $T$ and list of attributes $\textbf{Y}$, let ${\sf MB}_{k, i}$ denote a MB with best tuple $s_{k, i}$ among all MBs of length $k$ in $T[i]$.
If 
    $s_{k, i} \preceq_{\Delta, \textbf{Y} \upA} t_{i+1}$,
then there are two candidates for ${\sf MB}_{k+1, i+1}$$:$ ${\sf MB}_{k+1, i}$ with $s_{k+1, i}$ being its maximal tuple; and a new ${\sf MB}_{k, i} \cup \{t_{i+1}\}$ with its maximal tuple over $s_{k, i}$ and $t_{i+1}$, i.e., $\max_{\textbf{Y}}(s_{k, i}, t_{i+1})$. 

\begin{enumerate}[nolistsep,leftmargin=*]  
  \item If $s_{k+1, i}$ is not a minimal tuple among $\{s_{k, i}, s_{k+1, i}, t_{i+1}\}$, then ${\sf MB}_{k+1, i+1} = {\sf MB}_{k, i} \cup \{t_{i+1}\}$ and $s_{k+1, i+1} = \max_{\textbf{Y}}(s_{k, i}, t_{i+1})$.
  \item Else, ${\sf MB}_{k+1, i+1} = {\sf MB}_{k+1, i}$ and $s_{k+1, i+1} = s_{k+1, i}$. \rbox
  \end{enumerate}
\end{theorem}

\eatproofs{
\begin{proof}
Consider first Theorem~\ref{them:best} case 1. Since $d \mbox{(} t_{i+1}.\textbf{Y}, s_{k, i}.\textbf{Y}  \mbox{)}$ $\leq \Delta$, tuple $\max_{\textbf{Y}}\{s_{k, i}, t_{i+1}\}$ is the maximal tuple of a new MB with length $k+1$: ${\sf MB}_{k, i} \cup t_{i+1}$. In addition,
$\min_{\textbf{Y}}\{s_{k+1, i}, \max_{\textbf{Y}}\{s_{k, i}, t_{i+1}\}\}$ $=$ $\max_{\textbf{Y}}\{s_{k, i}, t_{i+1}\}$, therefore, $\max_{\textbf{Y}}\{s_{k, i}, t_{i+1}\}$ is the best tuple among MBs with length $k+1$ in $T\mbox{[}i+1\mbox{]}$. Theorem~\ref{them:best} case 2 follows analogically. 
\end{proof}
}

Based on Lemma~\ref{pro:lmb}, best tuples for each possible $k, i \in \{1, \cdots, n\}$ can be computed recursively.

  \begin{lemma}\label{pro:lmb}
Let $s_{0, i}.\textbf{Y}$ equal to 0
  for
  $i \in \{ 0, \cdots, n \}$ and $s_{k, 0}.\textbf{Y}$ equal to $\infty$ 
  for $k \in \{ 1, \cdots, n \}$. A best tuple
  $s_{k+1, i+1}$ of a MB with length
  $k+1$ in a prefix $T[i+1]$ satisfies the following
  recurrence, where
  $u = \min_{\textbf{Y}} \mbox{(}s_{k+1, i},
  \max_\textbf{Y}\mbox{(}t_{i+1}, s_{k, i}\mbox{)}
  \mbox{)}$.
  \begin{align*}\label{eq:b_ik}\fontsize{9.5}{9.5}
    \begin{split}
      & s_{k+1, i+1} =
       \left \{
        \begin{array}{rl}
          u & \text{if } 
          s_{k, i} \preceq_{\Delta, \textbf{Y} \upA} t_{i+1} \\
         s_{k+1, i} & \text{otherwise}
        \end{array}
      \right.		
    \end{split}
  \end{align*}\rbox
\end{lemma}

\begin{example}\label{eg:lib} 
  Consider $T=\{t_1-t_9\}$ over Table~\ref{tab:order}, $\textbf{Y} = [\sf{year}]$ and $\Delta =1$.
  Initially, 
  $s_{0, i}.\textbf{Y}$ is set to $0$ and
  $s_{k, 0}.\textbf{Y}$ to $\infty$ for
  $i \in \{ 0, \cdots, 9 \}$ and  $k \in \{ 1, \cdots, 9 \}$.
%
  First tuple $t_1$ (with {\sf year} $'92$) is checked, if it can extend any
  MB.
  Since 
     $s_{0, 0} \preceq_{\Delta, \textbf{Y} \upA} 92$
  a new MB of length one with a maximal tuple $t_1$ is found. $s_{1, 1}$ is set
  to $\max_\textbf{Y}\mbox{(}t_1.{\sf year}, s_{0, 0}\mbox{)}$ = $'92$.
  For each $k \in \{ 1, \cdots, 8 \}$, 
    $92 \preceq_{\Delta, \textbf{Y} \upA} s_{k, 0}$.
  Thus, $s_{k+1, 1}$
  is set to $s_{k+1, 0}$ = $\infty$.
 Tuples $\{ t_2$-$t_{9} \}$ are processed accordingly.
   \rbox
\end{example}

Also, Lemma~\ref{lemma:monotonicity_best_tuples} states that best tuples are ordered by lengths of their monotonic bands. 

\begin{lemma}\label{lemma:monotonicity_best_tuples}
For each $i \in [0, n]$, best tuples in $T[i]$ are monotonically ordered, i.e., $\forall_{k_1, k_2 \in [0, n], k_1 < k_2} s_{k_1, i} \preceq_{\textbf{Y} \upA} s_{k_2, i}$. \rbox
\end{lemma}

\eatproofs{
\begin{proof}
   In case of Theorem~\ref{them:best}.1, $s_{k+1, i+1} = \max_{\textbf{Y}}(s_{k, i}, t_{i+1})$, hence, $s_{k, i} \preceq_{\textbf{Y} \upA} s_{k+1, i+1}$; In case of Theorem~\ref{them:best}.2, given $s_{k+1, i+1} = s_{k+1, i}$, we know $s_{k, i} \preceq_{\textbf{Y} \upA} s_{k+1, i+1}$, therefore, $s_{k, i} \preceq_{\textbf{Y} \upA} s_{k+1, i+1} = s_{k+1, i}$.
\end{proof}
}


\eat{As shown in Figure~\ref{fig:db_table},} 

Thus, the exhaustive computation to store double-array of all best tuples can be avoided by calculating sorted single-array with best tuples of length $k$ in $T$ by keeping track of the shortest and longest length of MBs with a best tuple that ends at $t_{i}$ in $T[i]$. This is the crux of the solution to compute a LMB in $O(n \log n)$ time with Algorithm~\ref{alg:lmb} presenting the pseudo-code. 
%
%
Array $B_{\sf inc}$ stores a best tuple $s_{k}$ for each
$k \in \{1, \cdots, n \}$ over the sequence, i.e., $B_{\sf inc}\mbox{[}k\mbox{]} = s_k$. Initially,
  $B_{\sf inc}\mbox{[}k\mbox{]} = t_{\infty}$ with $t_{\infty}.\textbf{Y} =\infty$ for each $k \in \{1, \cdots, n \}$.
For each tuple $t_i$ in $T$, $B_{\sf inc}$ is updated by finding the {\em left-most position} of $t_i$ in
$B_{\sf inc}$, denoted by $k_1$ to $k_2$, as follows (Line~\ref{line:alg1_5}).

\begin{itemize}[nolistsep,leftmargin=*]
\itemsep=0pt
\item $k_1$ is the smallest index in $B_{\sf inc}$ that satisfies $t_i$ $\prec_{\textbf{Y} \upA}$ $s_{k_1}$.
It is the shortest length of MBs with a best tuple that ends at $t_i$ in $T\mbox{[}i\mbox{]}$. 
\item $k_2$ is the smallest index in $B_{\sf inc}$ that satisfies $t_i$  $\prec_{\Delta, \textbf{Y} \upA}$ $s_{k_2}$.
It is the longest length of MBs
  with a best tuple that ends at $t_i$ in $T\mbox{[}i\mbox{]}$.
\end{itemize}	   

$P_{\sf inc}$ is an array of size $n$ that
stores the set of lengths over shortest and longest MBs with best tuples ending at
$t_i$ for each $i \in \{1, \cdots, n \}$, i.e.,
$P_{\sf inc}\mbox{[}i\mbox{]}=\{k \mbox{ } | \mbox{ } k \in \{ k_1, \cdots,  k_2 \} \}$. For each
$k \in \{ k_1$, $\cdots$, $k_2 \} $, $B_{\sf inc}\mbox{[}k\mbox{]}$ is updated by $\max_\textbf{Y}\mbox{(}s_{k-1}, t_i\mbox{)}$ (Theorem~\ref{them:best}, case 1) and adding $\{ k_1$, $k_2 \}$ to $P_{\sf inc}\mbox{[}i\mbox{]}$ (Line~\ref{line:alg1_11}). Given band-width $\Delta$, each tuple $t_{i}$ can contribute to maximally $\Delta$ MBs, i.e., the sorted array can maximally be updated $\Delta$ times for each tuple. 

\begin{figure}[t]\centering
\paddingT
  \includegraphics[scale=0.5]{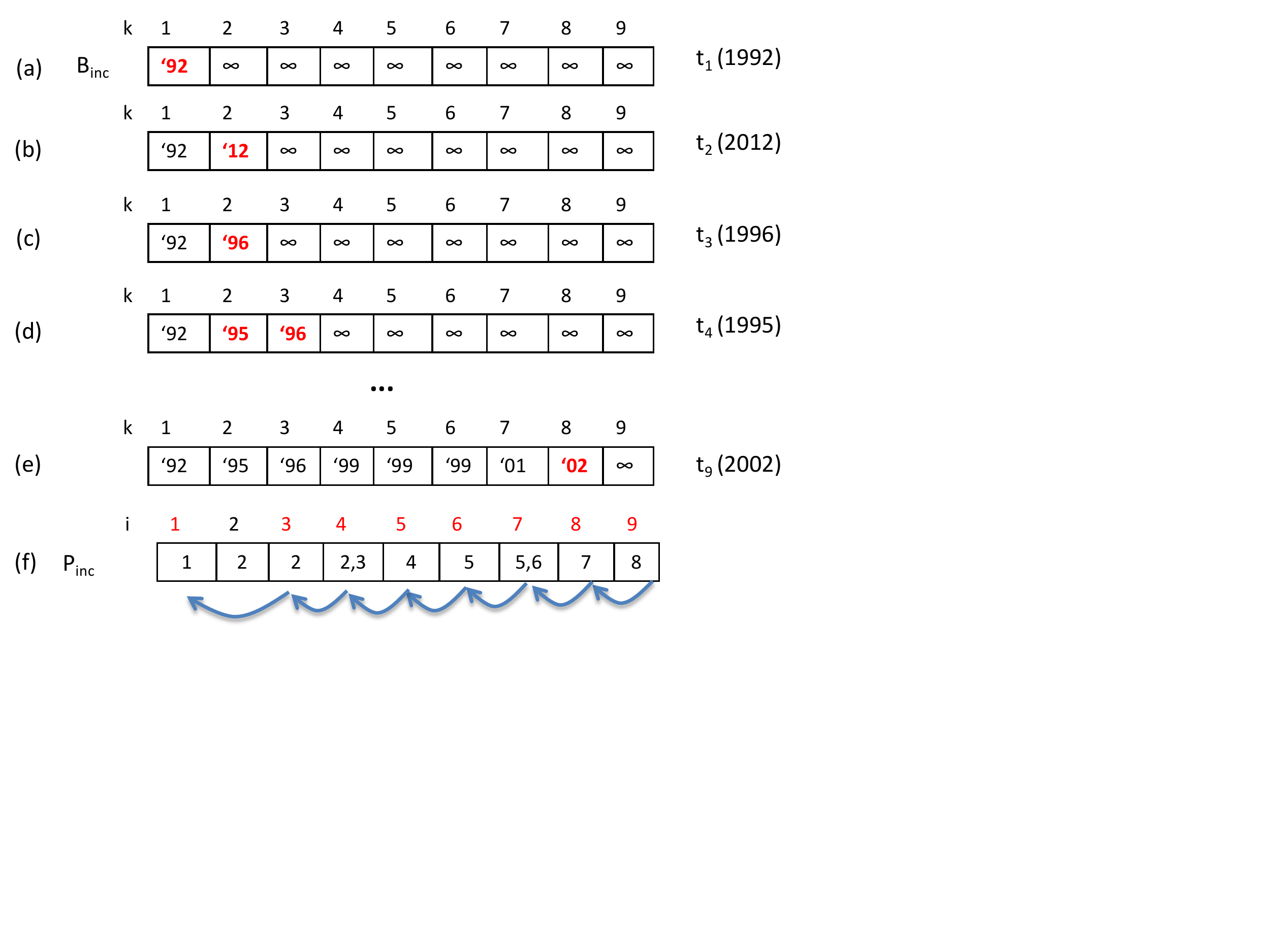}
 \vspace{-1.5in}
  \caption{\small{Finding LMB; tuples in $B_{\sf inc}$ are represented by year.}}\label{fig:dp_insertion}
    \paddingD
\end{figure}

\begin{example}\label{ex:dp_insertion}
Consider $T=\{t_1-t_9\}$ over Table~\ref{tab:order}, $\textbf{Y} = [\sf{year}]$ and $\Delta =1$. Since, initially 
    $B_{\sf inc}[k]$ =  $\infty$ and
    $t_1$ $\prec_{\textbf{Y} \upA}$ $s_{1}$ and $t_1$  $\prec_{\Delta, \textbf{Y} \upA}$ $s_{1}$, thus, $k_1 = k_2$ = $1$ and $B_{\sf inc}[1]$ = $\max_{\textbf{Y}}$($s_0, t_1$) = $'92$. Next, since $t_2$ $\prec_{\textbf{Y} \upA}$ $s_{2}$ and $t_2$  $\prec_{\Delta, \textbf{Y} \upA}$ $s_{2}$, hence, $k_1 = k_2$ = $2$ and $B_{\sf inc}[2]$ = $\max_{\textbf{Y}}$($s_1, t_2$) = $'12$. As $t_3$ $\prec_{\textbf{Y} \upA}$ $s_{2}$ and $t_3$  $\prec_{\Delta, \textbf{Y} \upA}$ $s_{2}$, thus, $k_1 = k_2$ = $2$ and $B_{\sf inc}[2]$ = $\max_{\textbf{Y}}$($s_1, t_3$) = $'96$. Then, since $t_4$ $\prec_{\textbf{Y} \upA}$ $s_{2}$ and $t_4$  $\prec_{\Delta, \textbf{Y} \upA}$ $s_{3}$, hence, $k_1 = 2$ and $k_2$ = $3$, $B_{\sf inc}[2]$ = $\max_{\textbf{Y}}$($s_1, t_4$) = $'95$ and $B_{\sf inc}[3]$ = $\max_{\textbf{Y}}$($s_2, t_4$) = $'96$.
The details of computing $B_{\sf inc}$ and $P_{\sf inc}$ for the following steps are reported in Fig.~\ref{fig:dp_insertion}.  
\rbox

%
\end{example}

  \begin{algorithm}[htbp]
    \caption{Computing LMB}
  \label{alg:lmb}%
	\SetKw{AAnd}{and}
	\SetKw{New}{new}
	\SetKwFunction{BinarySearch}{binary\_search}
	\SetKwFunction{PredecessorInc}{leftMostPosition}
	\SetKwFunction{PredecessorDec}{posDec}
	\SetKwFunction{Add}{append}
	\SetKwFunction{Max}{$\max_\textbf{Y}$}
	\SetKwFunction{Min}{$\min_\textbf{Y}$}
	\SetKwFunction{Contain}{contain}
	\SetKwFunction{Last}{last}	
	\SetKwFunction{Band}{band}
	\SetKwFunction{IsEmpty}{is\_empty}
  %
	\KwData {$T=\{t_1, t_2, \cdots, t_n\}$, band width $\Delta$}
  \KwResult{LMB  in $T$ }
  \BlankLine
	\For{$i \leftarrow 1$ \textbf{to} $n$}
	{
		$t_\infty.\textbf{Y} \leftarrow \infty$; $t_0.\textbf{Y} \leftarrow 0$; $B_{\sf inc}\mbox{[}i\mbox{]} \leftarrow t_\infty$\;
    $P_{\sf inc}\mbox{[}i\mbox{]} \leftarrow \emptyset$\;
	}
	$k_{\sf inc} = 0$\; 
	\For{$i \leftarrow 1$ \textbf{to} $n$}
	{
		$k_1 \leftarrow\,$\PredecessorInc{$B_{inc}, t_i, 0$};
    $k_2 \leftarrow\,$\PredecessorInc{$B_{\sf inc}, t_i, \Delta$} \;\label{line:alg1_5}
		$k_{\sf inc} \leftarrow$ \Max{$k_{\sf inc}, k_2$} \;\label{line:alg1_7}
		\For{$k \leftarrow k_2$ \textbf{to} $k_1$}
		{
			$b \leftarrow t_0$ \;
			\lIf{$k > 1$}
			{
				$b \leftarrow B_{\sf inc}\mbox{[}k-1\mbox{]}$
			}
			$B_{\sf inc}\mbox{[}k\mbox{]} \leftarrow$ \Max{$b, t_i$};  \Add{$P_{\sf inc}\mbox{\emph{[}}i\mbox{\emph{]}}, k$}\;\label{line:alg1_11}
		}
	}
	$l \leftarrow$ the maximal value in $P_{\sf inc}$\; 
	\For{$i \leftarrow n$ \textbf{to} $1$\label{line:alg1_17}}
	{
	    \If{$P_{\sf inc}[i]$ contains $l$}
	    {
	        ${\sf LMB}[l] \leftarrow t_{i}$\;
	        $l \leftarrow l-1$\label{line:alg1_20}\;
	    }
	}
   \Return $LMB$ \;
\end{algorithm}

Next, we describe how to compute a LMB. The path of a LMB is constructed in a sequence of tuples $T$ in reverse order by scanning the array $P_{\sf inc}$.  Let $k \in P_{\sf inc}\mbox{[}i_k\mbox{]}$ be the largest value in $P_{\sf inc}$, i.e., there exists a LMB of the longest length $k$ in a sequence of tuples $T$; \eat{A LMB in  $T$ with length $k$} and $t_{i_k}$ is found as the $k^{th}$ tuple in the LMB.  Starting from $P_{\sf inc}\mbox{[}i_k\mbox{]}$ and $k$, $P_{\sf inc}$ is scanned in reverse order until the first tuple $P_{\sf inc}\mbox{[}i_{k-1}\mbox{]}$ is found that contains $k-1$. Then, $t_{i_{k-1}}$ is found, the $k-1^{th}$ tuple in the LMB. $P_{\sf inc}$ is continued to be scanned until all $k$ tuples in the LMB are found (Lines~\ref{line:alg1_17}--~\ref{line:alg1_20}).

\begin{example}\label{ex:lmb_t1_t9}
  Continuing Example~\ref{ex:dp_insertion}, 
  Fig.~\ref{fig:dp_insertion}.(f) reports
  the array $P_{\sf inc}$  for finding a LMB.
 $P_{\sf inc}$ is scanned to find the largest
  value $8$ in $P_{\sf inc}[9]$. Thus, a LMB with length 8 exists in $T$ and $t_{9}$ is its
  eighth tuple.  By a reverse scan (marked with
  arrows in Fig~\ref{fig:dp_insertion}) from $P_{\sf inc}[9]$, the $7$-th tuple $t_{8}$ is found. The operation is
  continued until all tuples in a LMB are found; i.e., 
  $\{t_1 ('92)$, $t_3 ('96)$, $t_4 ('95)$, $t_5 ('99)$, $t_6 ('00)$, $t_7
  ('99)$, $t_8 ('01)$, $t_{9} ('02)\}$. \rbox
%

  
\end{example}

\begin{theorem}\label{lemma:lmb}
  Algorithm~\ref{alg:lmb} correctly finds a LMB in a sequence of tuples $T$ of size $n$ in
  $O\mbox{(}n \log n\mbox{)}$ time and $O\mbox{((}\Delta+1\mbox{)} n\mbox{)}$ space.\rbox
\end{theorem}

\eatproofs{
\begin{proof}
   To find a LMB in the sequence $T$, best tuples are the key. Since a tuple $B_{\sf inc}\mbox{[}k_1\mbox{]}$ is updated by $\max\mbox{(}s_{k_1-1}.\textbf{Y}, t_i.\textbf{Y}\mbox{)}$ accordingly to Algorithm~\ref{alg:lmb}, where 
   $t_i$ $\prec_{\textbf{Y} \upA}$ $s_{k_1}$, the corresponding band $\sf{MB}_{k_1, i}$ is a MB with smallest maximal tuples that ends at tuple $t_i$ in the sequence $T\mbox{[}i\mbox{]}$. It is also a monotonic band with the shortest length, as $k_1$ is the smallest index in $B_{\sf inc}$. Similarly, $\sf{MB}_{k_2, i}$ is an {\sf MB} of the longest length among MBs with the smallest maximal tuple that ends at $t_i$ in $T\mbox{[}i\mbox{]}$.
For each $t_i \in T$, the lengths of MBs with the smallest
  maximal tuples that end at $t_i$ fall into range
  $\mbox{[}k_1, k_2\mbox{]}$. Therefore, the length of a LMB in
  $T\mbox{[}i\mbox{]}$ is the maximal value in array
  $P_{\sf inc}\mbox{[}i\mbox{]}$.

For each tuple in $T$ of size $n$, it takes $O\mbox{(}\log n\mbox{)}$ time  to update arrays $B_{\sf inc}$ and $P_{\sf inc}$, since they are maintained sorted. Thus, Algorithm~\ref{alg:lmb} takes $O\mbox{(}n\log n\mbox{)}$ time to find a $\sf LMB$ in $T$ . For each tuple $t_i$ maximally $\Delta + 1$ values are inserted into array $P_{\sf inc}$. Thus, Algorithm~\ref{alg:lmb} takes $O\mbox{((}\Delta + 1\mbox{)}n\mbox{)}$ space.
\end{proof}
}

\subsection{Band-Width and Candidate Dependencies} 
\label{sub:estimating_parameters}
Our goal is to effectively identify outliers in a
sequence of tuples, while being tolerant to tuples that slightly violate a traditional OD. Since band ODs may hold over subsets of data called series, to identify the correct band-width, we partition the entire sequence of tuples $T$ (ordered by \textbf{X}) over a table $r$ into contiguous subsequences of tuples $S$. We identify contiguous subsequences of tuples by using divide-and-conquer method, such that tuples in $S$ satisfy a traditional OD $\textbf{X}$ $\mapsto$ $\textbf{Y}$ within approximation ratio. 

We would like to include a large number of tuples from each sequence $S$ into a LMB by setting an optimal \emph{band-width} $\Delta$, such that the {\em distances} of outliers from a LMB are {\em large}. To capture this, we propose a method to \emph{automatically} compute the optimal band-width based on LMBs. 
For a particular band-width $\Delta$, $d_{\Delta}$ denotes a {\em distance} of outliers from a LMB and $a_{\Delta}$ denotes a {\em distinctive degree} of $\Delta$ in a sequence of tuples $S$.

\begin{equation}\label{eq:delta_t}
  a_{\Delta} = \left \{
    \begin{array}{rl}
      0	  \mbox{ if } \Delta  = 0; 
      \mbox{ } \mbox{ }  \mbox{ }  \mbox{ }  \mbox{ }  \mbox{ }  \mbox{ }    
      \frac{d_{\Delta} - d_{\Delta - 1} }{d_{\Delta }}
          \text{ otherwise}.
    \end{array} \right.	
\end{equation}

\eat{
For each outlier tuple $t_{j}$ in $S = \{ t_{1}, \cdots, t_{j}, \cdots, t_{n}  \}$, let $t_{j}'.\textbf{Y}$ denote a repair of $t_{j}.\textbf{Y}$. If $S$ is a sequence where LMB is a LMB (LDB), then $t_{left}$ denotes the maximal (minimal) tuple in $T[j-1]$ that is part of a LMB (LDB) in $S$\eat{$\max ( t_{1}.\textbf{Y}, \cdots, t_{j-1}.\textbf{Y} )$}; and $t_{right}$ denotes the minimal (maximal) tuple in $T[j+1,n]$ that is part of a LMB (LDB)\eat{$\min ( t_{j+1}.\textbf{Y}, \cdots, t_{n}.\textbf{Y} )$}.\eat{ Accordingly, if $S$ is a sequence where LMB is a LDB, then $t_{left}.\textbf{Y}$ denotes \colB{$\min ( t_{1}.\textbf{Y}, \cdots, t_{j-1}.\textbf{Y} )$} and $t_{right}.\textbf{Y}$ denotes $\max ( t_{j+1}.\textbf{Y}, \cdots, t_{n}.\textbf{Y} )$.} We define the estimated repair $t_{j}'.\textbf{Y}$ as $(t_{left}.\textbf{Y}$ $+$ $t_{right}.\textbf{Y})$ $/$ $2$.}

For each outlier over a tuple $t_{j}$ in $S = \{ t_{1}, \cdots, t_{j}, \cdots, t_{n}  \}$, let $t_{j}'.\textbf{Y}$ denote a repair of $t_{j}.\textbf{Y}$. Let $t_{left}$ denotes a maximal tuple in $T[j-1]$ that is part of a LMB in $S$\eat{$\max ( t_{1}.\textbf{Y}, \cdots, t_{j-1}.\textbf{Y} )$}; and $t_{right}$ denotes a minimal tuple in $T[j+1,n]$ that is part of a LMB \eat{$\min ( t_{j+1}.\textbf{Y}, \cdots, t_{n}.\textbf{Y} )$}.\eat{ Accordingly, if $S$ is a sequence where LMB is a LDB, then $t_{left}.\textbf{Y}$ denotes \colB{$\min ( t_{1}.\textbf{Y}, \cdots, t_{j-1}.\textbf{Y} )$} and $t_{right}.\textbf{Y}$ denotes $\max ( t_{j+1}.\textbf{Y}, \cdots, t_{n}.\textbf{Y} )$.} We define the estimated repair $t_{j}'.\textbf{Y}$ as $(t_{left}.\textbf{Y}$ $+$ $t_{right}.\textbf{Y})$ $/$ $2$.

\begin{example}
Consider $S=\{t_1-t_9\}$ in Table~\ref{tab:order} and let $\Delta =1$. Since the value 2012 of the tuple $t_{2}$ over attribute $\sf{year}$ is incorrect\eat{ and $t_
{2}$ is part of a LMB wrt $\sf{year}$}, the repair $t_{2}'.\sf{year}$ is calculated as (1992 + 1995) / 2, which is rounded to 1993. \rbox
\end{example}


The distance $d\mbox{(}t, \sf{LMB}\mbox{)}$ of tuple $t$ from a LMB is computed as $|d\mbox{(}t'.\textbf{Y}, t.\textbf{Y}\mbox{)}|$.  
%
The distance $d_{\Delta}$ of outliers from a LMB is calculated as the average distance
i.e., $d_{\Delta}$ = $\sum_{t \notin {\sf LMB}, t \in S}{d\mbox{(}t, {\sf LMB}}\mbox{)}\ $/\ $|\{t: t
\notin {\sf LMB}, t \in S\}|$.
%
The band-width $\Delta$ is chosen that \emph{maximizes the distinctive degree} $a_{\Delta}$. 
Note that since entire sequence $T$ is divided into contiguous
subsequences $S$, the band-width $\Delta$ is the average
  aggregated value computed over all subsequences $S$.


To identify candidate abcODs without human intervention, we use a global approach to find all traditional ODs within an approximation ratio~\cite{SGG18,DBLP:journals/pvldb/SzlichtaGGKS17} to narrow the search space, as discovering traditional ODs is less computationally intensive. Since band ODs may hold over subsets of the data (with a mix of ascending and descending ordering), we separate an entire sequence of tuples into contiguous subsequences of tuples by using \emph{divide-and-conquer} approach, such that tuples over contiguous subsequences satisfy a traditional OD within approximation ratio. Found traditional ODs ranked by the measure of interestingness~\cite{SGG18,DBLP:journals/pvldb/SzlichtaGGKS17} are used as candidate embedded band ODs for the abOD, bcOD and abcOD discovery problems (Sections~\ref{sub:abOD_discovery}--\ref{sec:bidDiscovery}).

\begin{example}
\label{ex:lmbpinc}
   Assume band-width is computed for an attribute
    $\sf{year}$ over Table~\ref{tab:order} wrt an OD between
    $\sf{cat\#}$ and $\sf{year}$ and an approximation ratio of 0.4 (set
    higher for traditional ODs as they do not take band-width into
    account). The divide-and-conquer method 
    divides Table~\ref{tab:order} into $T_1 = \{
    t_{1}$--$t_{5} \}$, $T2 = \{ t_{6}$--$t_{10} \}$,
    $T3 = \{ t_{11}$--$t_{16} \}$ and $T4 = \{ t_{17}$--$t_{22}
    \}$. Since distinctive degree value is the highest for band-width
    of 1 over $T_1$ and $T_4$, 2 over $T_2$ and 0 over $T_3$ the averaged band-width $\Delta = 1$ $($rounded from $(1 + 2 + 0 + 1)
    / 4))$.
\rbox
\end{example}

\subsection{\MakeLowercase{ab}OD\MakeLowercase{s} Discovery}\label{sub:abOD_discovery}

In practice, band ODs may not hold exactly, due to errors in the data, but approximately with some exceptions. 
Given a band OD $\textbf{X} \mapsto_{\Delta} \overline{\textbf{Y}}$, 
the goal is to verify whether a band OD holds, such that inconsistent
tuples that severely violate a band OD are few. 
As in prior work on functional dependency discovery~\cite{DBLP:journals/cj/HuhtalaKPT99}, the minimum number of tuples are computed that must be removed from the given table for the band OD to hold.
The problem of discovering approximate band ODs is defined as follows~\cite{LSBS20}. 

\begin{definition}[abOD Discovery]
\label{def:series}
Given a band OD $\varphi$: $\textbf{X} \mapsto_{\Delta } \overline{\textbf{Y}}$ and 
table $r$,
the approximate band OD discovery problem is to 
identify the minimal set of tuples that violate a band OD  with an error ratio
$e(\varphi)$ = $min \{|t| \mbox{ } : \mbox{ } t \subseteq r, r \setminus t \models \varphi\}$$/$$|r|$.
\rbox
 \end{definition}
 
The measure $e(\varphi)$ has a natural interpretation as the fraction of tuples with inconsistencies affecting the dependency. Band ODs that hold approximately with some exceptions are called \emph{approximate band ODs} (abODs).
The \emph{approximate band OD} discovery problem can be solved by finding a LMB in a sequence of tuples $T$ over a table $r$ ordered by $\textbf{X}$. The minimal set of tuples that violate a band OD $\textbf{X} \mapsto_{\Delta } \overline{\textbf{Y}}$ are inconsistent tuples $s$, such that $s \in $T$, s \notin \sf {LMB}$.  

\begin{lemma}
\label{lemma:discabOD}
The approximate band OD discovery problem is solvable by finding a LMB with an error ratio $e(\varphi)$ $=$ $|s \notin \sf {LMB}|$ $/$ $|s \in \sf T|$.
\end{lemma}
  
\eatproofs{
\proof{
The proof follows directly from the definition of LMBs (Definition~\ref{def:2}).}
}


  \begin{example}
  Consider the approximate band OD $\varphi$: $\sf{cat\#}$ $\mapsto_{\Delta=1}$ $\sf{year} \upA$ over a table with tuples $\{t_1$--$t_9 \}$ in Table~\ref{tab:order}.
  Given Example~\ref{ex:lmbpinc}, a LMB of length $8$ is found, i.e., $\{t_1 ('92)$, $t_3 ('96)$, $t_4 ('95)$, $t_5 ('99)$, $t_6 ('00)$, $t_7
  ('99)$, $t_8 ('01)$, $t_{9} ('02)\}$. Thus, an error ratio $e(\varphi)$ = 1/9.
\end{example}

Since we provided a new optimization to the LMB computation (Theorem~\ref{lemma:lmb}), based on Lemma~\ref{lemma:discabOD}, abOD discovery is improved from $O(n^2)$ (as in prior work~\cite{LSBS20}) to $O(n \log n)$.

\begin{theorem}\label{theo:optAB}
  The abOD discovery problem can be solved in $O\mbox{(}n \log n\mbox{)}$ time, where $n$ is the number of tuples over a table. \rbox
\end{theorem}

%% file: conditionalBand.tex
\eat{
The optimal solution ${\sf OPT}\mbox{(}j\mbox{)}, j \in \{ 1, \cdots, n \}$ in a prefix $T\mbox{[}j\mbox{]}$ contains optimal solutions to the subproblems in prefixes $T\mbox{[}1\mbox{]}, T\mbox{[}2\mbox{]}, \cdots, T\mbox{[}j-1\mbox{]}$. 
\begin{equation}\label{eq:opt}
\fontsize{9.5}{9.5}
  {\sf OPT}\mbox{(}j\mbox{)} = \left \{
    \begin{array}{ll}
      0 &  j = 0  \\[10pt]
      \min_{i \in \{1, \cdots, j-1 \}}\{{\sf OPT}\mbox{(}i\mbox{)} + g\mbox{(}T\mbox{[}i+1, j\mbox{]}\mbox{)} \} & j > 0
    \end{array} \right.
\end{equation}	

\subsection{Computing bcODs}
\label{sub:pieces}}

\eat{
\begin{figure*}[t]\centering
  \includegraphics[scale=0.35]{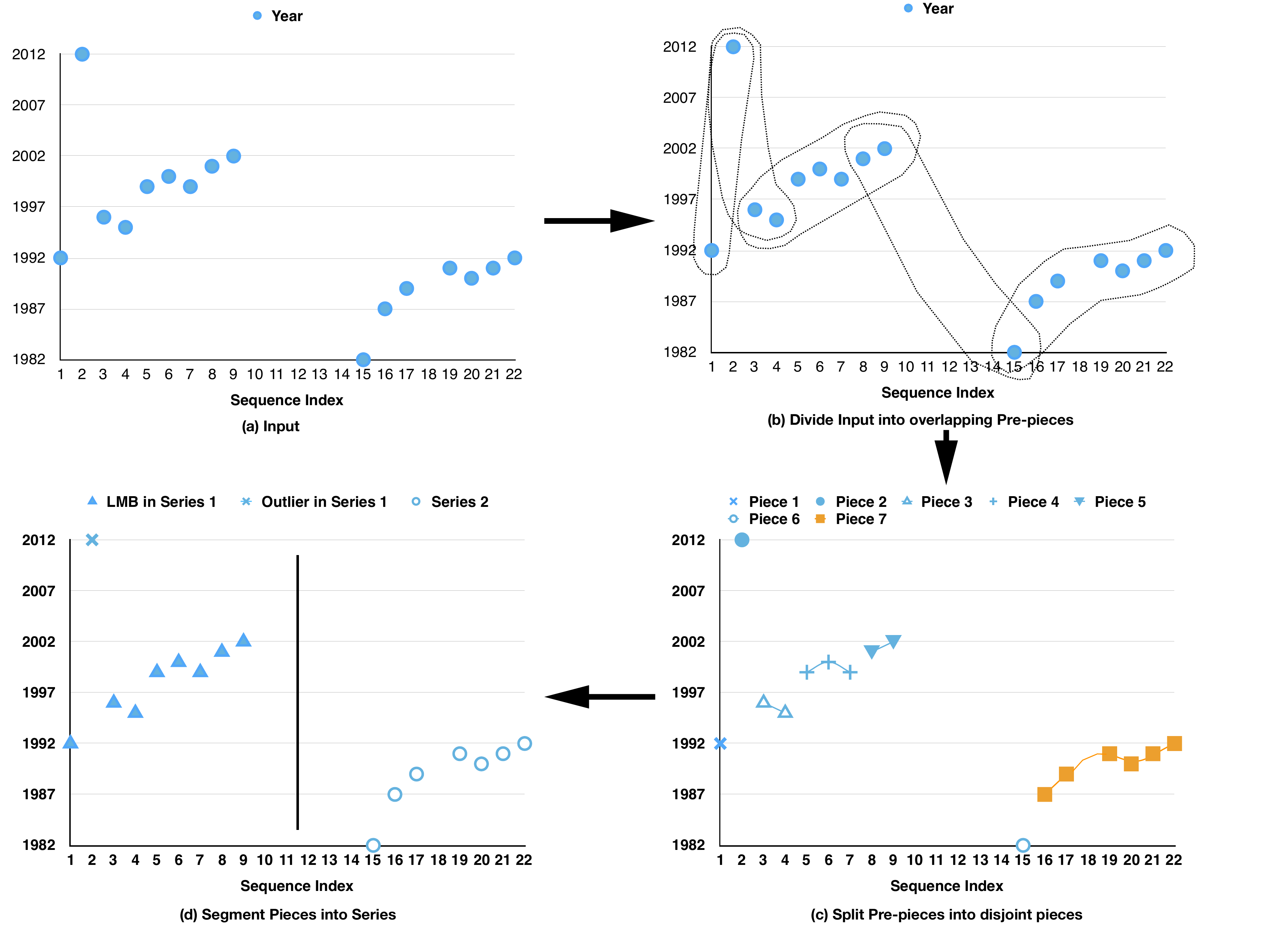}
    \paddingT
  \caption{\small{Using Pieces to compute abcODs.
    \label{fig:piece}}}
    \paddingD
\end{figure*}
}

\begin{figure*}[t]
\begin{multicols}{3}
    \includegraphics[scale=0.38]{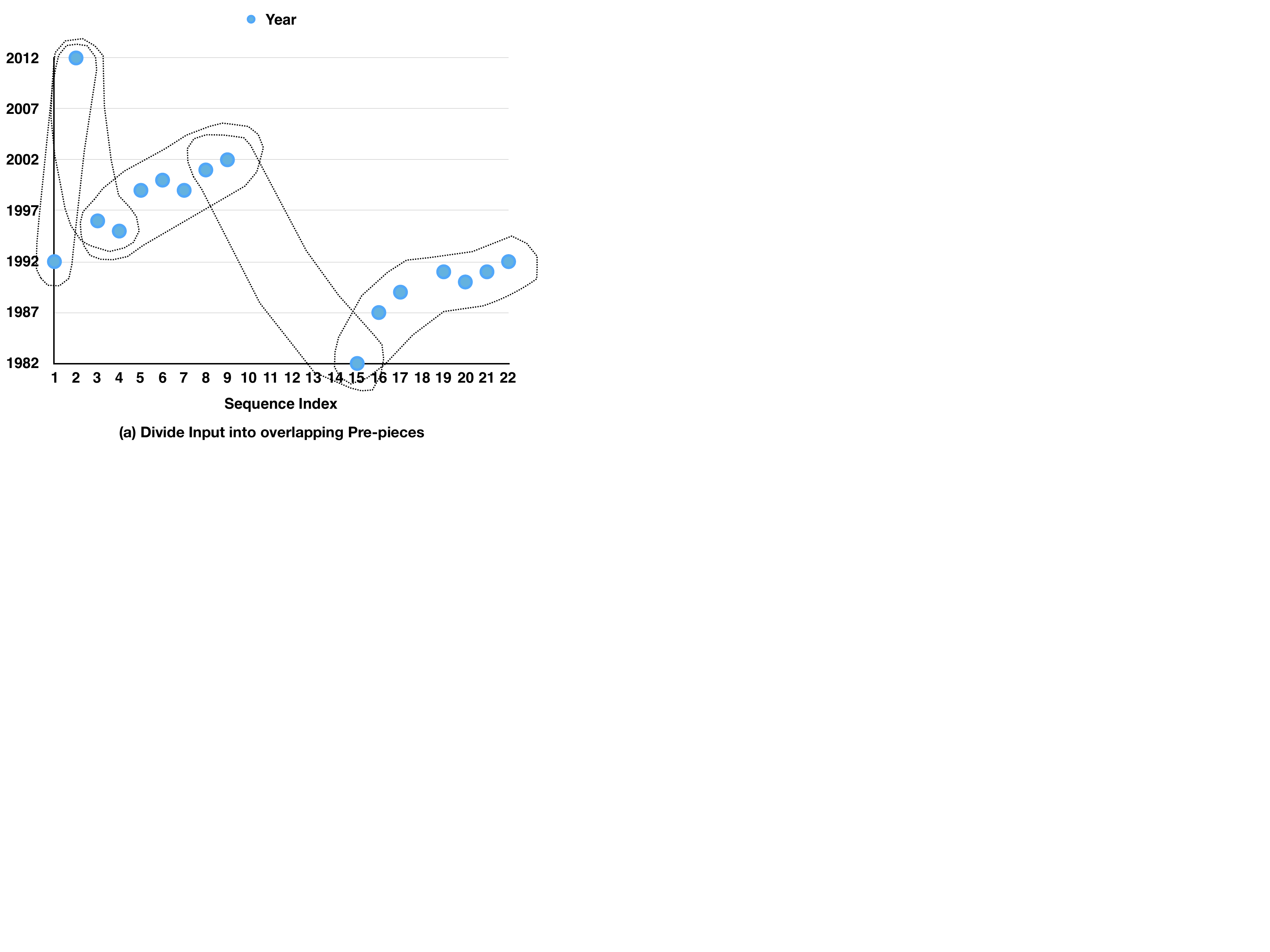}\par 
    \includegraphics[scale=0.38]{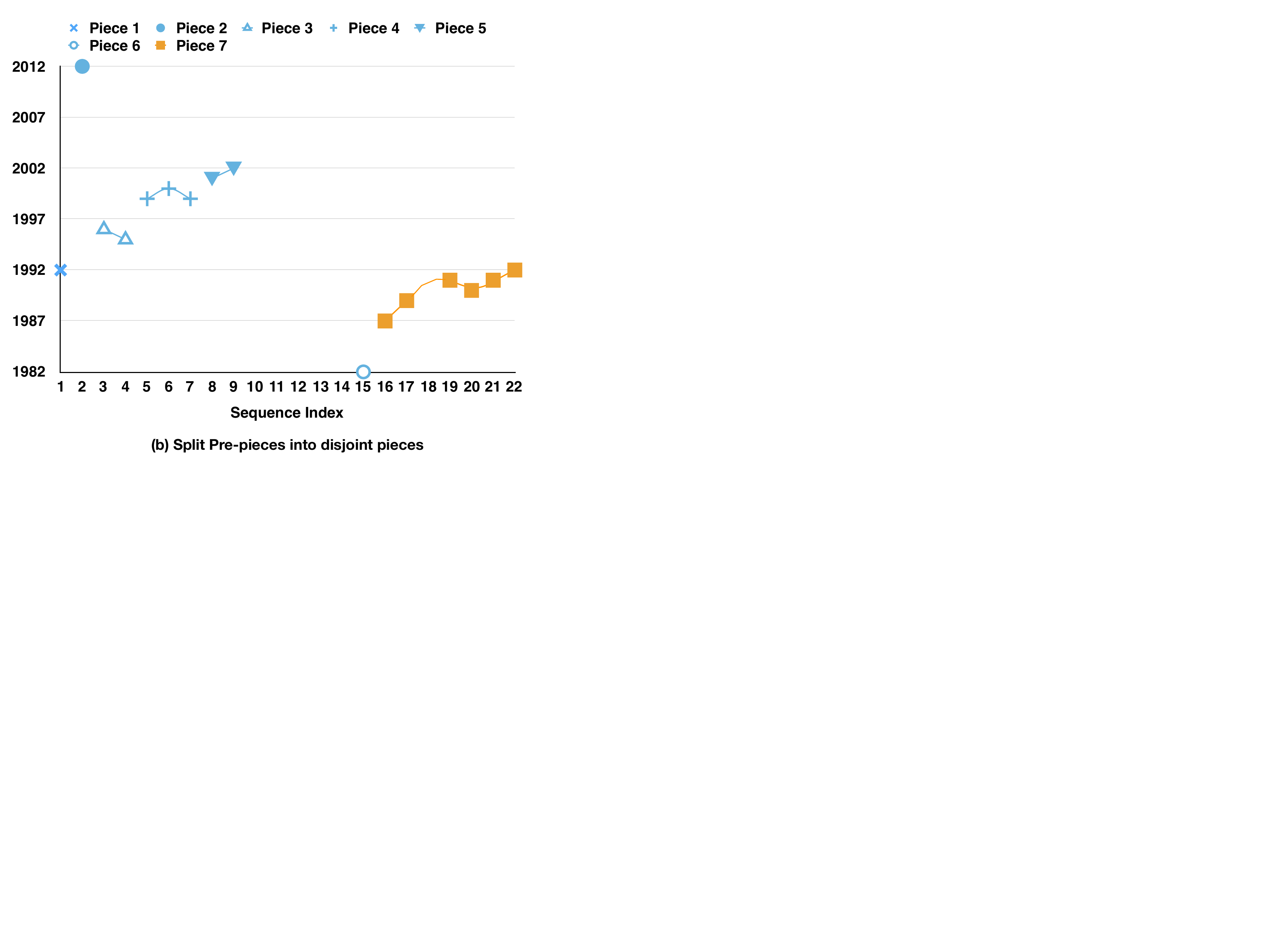}\par 
    \includegraphics[scale=0.38]{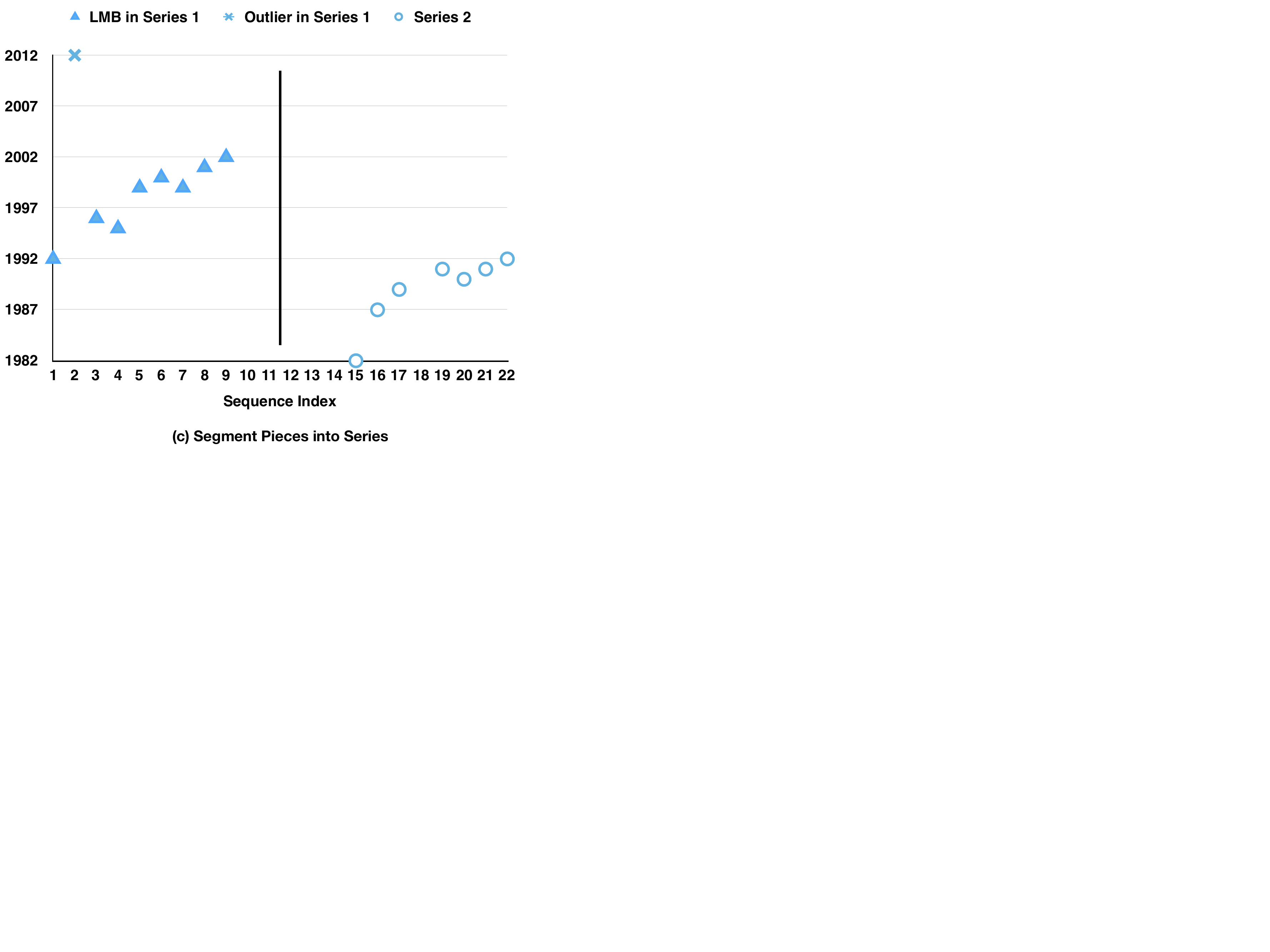}\par
    \end{multicols}
    \vspace{-5.8cm}
    \paddingT
\caption{\small{Using Pieces to compute abcODs.
    \label{fig:abcOD_with_piece}}}
    \paddingD
\end{figure*}

%% file: discovery.tex
\section{Discovery of \lowercase{abc}OD\lowercase{s}}
\label{sec:series}

To better understand the hierarchy of abcODs, initially, we consider the scenario when band ODs hold over subset of the data in Section~\ref{sec:bcODDisc} (bcODs). Both approximation and conditioning are studied in Sections~\ref{sub:data_segmentation_into_series}--~\ref{sub:pieces} (abcODs). 


\subsection{Discovery of \lowercase{bc}OD\lowercase{s}}
\label{sec:bcODDisc}


\eat{Given a band OD $\textbf{X} \mapsto_{\Delta} \textbf{Y}$, where $T$ is a sequence of tuples ordered by $\textbf{X}$, our goal is to find a minimal number of segments in $T$ such that a band OD holds on each segment.}


\eat{
\begin{definition}[Description Complexity]
Let $\textbf{X} \mapsto_{\Delta } {\textbf{Y}}$ be a band OD, $T$ be a sequence of tuples, ordered by $\textbf{X}$ over a table $r$. Let $S$ be a non-overlapping segmentation that splits $T$ into $m$ segments where $\textbf{X} \mapsto_{\Delta } {\textbf{Y}}$ holds in each\eat{i.e., $S = \{T[1, i], T[i+1, i+k], \cdots, T[i+j, n]\}$, $1 \leq i \leq n, 1 < k \leq j \leq n$}. The description complexity of segment $S_i$ with length $k$ is $\ln(k+1)$. The description complexity $g(S)$ of segmentation $S$ is $\sum_{i = 1}^m \ln(|S_i|)$, where $|S_i|$ is the length of segment $S_i$ in $T$. \rbox
\end{definition}

\begin{example}
Consider tuples $t_1-t_9$ in Table~\ref{tab:order} and a band OD $\sf{cat\#} \mapsto_{\Delta=1} \overline{\sf{year}}$. $S=\{S_1(t_{1-2}), S_2(t_{3-9})\}$ is a segmentation of $t_1-t_9$ where band OD $\sf{cat\#} \mapsto_{\Delta=1}$ holds on each. The description complexity of $S$ is computed as $\ln(2+1) + \ln(7+1) = 4.58$. \rbox
\end{example}
}

The band conditional OD discovery problem (without considering approximation) is defined as follows.

\begin{definition}[bcOD Discovery]
  Let $\textbf{X} \mapsto_{\Delta } {\overline{\textbf{Y}}}$ be a band OD, $T$ be a sequence of tuples ordered by $\textbf{X}$ over a table $r$. 
  Let $\bf\mathbb{S}$ denote all possible contiguous non-overlapping segmentations of $T$.
  %
  The \emph{band conditional OD (bcOD) discovery problem} is to find the segmentation amongst $\dot S \in \mathbb{S}$ with minimal number of segments, where a band OD $\textbf{X} \mapsto_{\Delta } {\overline{\textbf{Y}}}$ holds in each segment.  \eat{
  \begin{equation}\label{eq:bcod_opt}
    \begin{matrix}
      \displaystyle \min_{\dot S \in {\bf\mathbb{S}}} & g\mbox{(}\dot S\mbox{)}
    \end{matrix}
  \end{equation}}
  \rbox
\end{definition}

The bcOD discovery problem is solvable by scanning the sequence of tuples and splitting it into contiguous segments, 
whenever an {\em outlier} tuple appears; i.e., the first tuple that violates a band OD with respect to the current segment represented by its best tuple. The violating tuple becomes the first tuple in the next segment. 

\begin{theorem}\label{theorem:bcod}
  The bcOD discovery problem is solvable in $O(n)$ time and $O(n)$ space over a sequence of tuples $T$ of size $n$
  .\rbox
\end{theorem}

\eat{
\begin{proof}
TBA
\end{proof}
}

\eatproofs{
  \begin{algorithm}[htbp]
    \caption{Computing bcOD}
  \label{alg:bcod}%
	\SetKw{AAnd}{and}
	\SetKw{New}{new}
	\SetKwFunction{BinarySearch}{binary\_search}
	\SetKwFunction{PredecessorInc}{posInc}
	\SetKwFunction{PredecessorDec}{posDec}
	\SetKwFunction{Add}{append}
	\SetKwFunction{Max}{$\max_\textbf{Y}$}
	\SetKwFunction{Min}{$\min_\textbf{Y}$}
	\SetKwFunction{Contain}{contain}
	\SetKwFunction{Last}{last}	
	\SetKwFunction{Band}{band}
	\SetKwFunction{IsEmpty}{is\_empty}
  %
	\KwData {$T=\{t_1, t_2, \cdots, t_n\}$, band width $\Delta$}
  \KwResult{Segmentation $\dot S$ in $T$ }
  \BlankLine
	$\dot S \leftarrow \emptyset$; $\sf{best} \leftarrow t_{1}$; $S \leftarrow$ $\{t_1\}$\;
	\For{$i \leftarrow 2$ \textbf{to} $n$}
	{
	\eIf{$d(t_i.\textbf{Y}, \sf{best}.\textbf{Y}) \leq \Delta$}{
	    add $t_i$ to $S$;
	    $\sf{best} \leftarrow \max_{\textbf{Y}}(t_i, \sf{best})$
	}{
	add segment $S$ to $\dot S$\; 
	$S \leftarrow \{t_i\}$; $\sf{best} \leftarrow t_i$\;
	}
	}
   \Return $\dot S$ \;
\end{algorithm}
}

\begin{example}
  Consider a band OD ${\sf cat\#}$ $\mapsto_{\Delta = 1}$ ${{\sf year} \upA}$ $T$ $=$ $\{t_1-t_9\}$
  in Table~\ref{tab:order} ordered by ${\sf cat\#}$. The bcOD discovery problem solution consists of two segments: $\{t_1, t_2\}$ and $\{t_3$--$t_9\}$. 
  \rbox
\end{example}

Whereas the discovery of bcODs is relatively straightforward, there exist codependencies between abODs and bcODs that make the problem of abcODs discovery challenging (Sections~\ref{sub:data_segmentation_into_series}--~\ref{sub:pieces})

\subsection{Discovery of abcODs Problem}
\label{sub:data_segmentation_into_series}

To make band ODs relevant to real-world applications, we make them less strict to hold both \emph{approximately} with some exceptions and \emph{conditionally} on subsets of the data. We call these data dependencies \emph{approximate band conditional ODs} (abcODs). Specifically, given a band OD $\textbf{X} \mapsto_{\Delta} {\overline{\textbf{Y}}}$, where $T$ is a sequence of tuples ordered by $\textbf{X}$, our goal is to segment $T$ into multiple contiguous, non-overlapping subsequences of tuples, called {\em series}, such that (1) large fraction of tuples in each series satisfy a band OD (gain), and (2) outlier
tuples that severely violate a band OD in each series are few and sparse (cost). 
We experimentally verified in Sec.~\ref{sec:experimental_evaluation} that in practice errors are few and sparse over real-world datasets.


\begin{definition}[Gain]
Let $\textbf{X} \mapsto_{\Delta } {\overline{\textbf{Y}}}$ be a band OD, $T$ be a sequence of tuples ordered by $\textbf{X}$ over a table $r$. $|T_{nn}|$ denotes the number of non-null tuples in $T$, and $L_T$ denotes a LMB in $T$. The gain
$g(T)$ is defined as the portions of $T$ satisfying $\textbf{X} \mapsto_{\Delta } { \overline{\textbf{Y}} }$.
%
\begin{align}
\label{eq:gain_of_a_sequence}
      g(T) =
      | t : t &\in L_T,  t \in T| - |t: t \notin L_T,  t \neq null, t \in T|
  \end{align}
\end{definition}

\begin{example}\label{ex:gain_of_a_sequence}
Following Example~\ref{ex:lmb_t1_t9}, where the LMB $\{t_1$, $t_3-t_9\}$ of length $8$ is found, and $t_2$ is the only outlier, the gain 
is computed as $8-1=7$. \rbox
\end{example}

The outliers are penalized with the following cost function.

\begin{definition}[Cost]
Let $\textbf{X} \mapsto_{\Delta } {\overline{\textbf{Y}}}$ be a band OD, $T$ be a sequence of tuples ordered by $\textbf{X}$ over a table $r$. The cost 
$e(T)$ is the maximum number of contiguous
  outliers that violate $\textbf{X} \mapsto_{\Delta} {\overline{\textbf{Y}}}$ in $T$.
  \begin{equation}~\label{eq:error_of_a_sequence}
    \begin{matrix}
      e(T) = \displaystyle \max_{k \in \{ i, \cdots, j \}, 1 \leq i \leq j \leq |T|}  j + 1
      -i  :  \ t_k \notin L_T, t_k \in T
    \end{matrix}
  \end{equation}
\end{definition}

\begin{example}\label{ex:cost_of_a_sequdence}
Continuing with Example~\ref{ex:gain_of_a_sequence}. Since $t_2$ is the only outlier in sequence $\{t_1-t_9\}$, the cost 
is $1$. \rbox
\end{example}



We define the \emph{abcOD discovery problem} as a constrained optimization problem, aiming to maximize the gain function under the cost constraint. 

\begin{definition}[abcOD Discovery Problem]\label{def:series}
{
Let $\textbf{X} \mapsto_{\Delta } {\overline{\textbf{Y}}}$ be a band OD, $T$ be a sequence of tuples ordered by $\textbf{X}$ over a table $r$ and $\epsilon$ be an approximation error rate parameter. Let $\bf\mathbb{S}$ denote all possible contiguous, non-overlapping segmentations of $T$.
The \emph{approximate band conditional OD (abcOD) discovery problem} is to find the optimal segmentation among $\dot S \in \mathbb{S}$ that satisfies the following optimization function. 

  \begin{equation}\label{eq:series_opt}
    \begin{matrix}
      \displaystyle \max_{\dot S \in {\bf\mathbb{S}}} & g\mbox{(}\dot S\mbox{)}
      \qquad \textrm{s.t.}\  e\mbox{(}\dot S\mbox{)} \leq \epsilon
    \end{matrix}
  \end{equation}
  \eat{$g\mbox{(}\dot S\mbox{)}$ defines the \emph{gain} in terms of portions of $\dot S$ satisfying a band OD, and $e\mbox{(}\dot S\mbox{)}$ defines a \emph{cost} quantifying the number of \emph{errors} in $\dot S$ that violate a band OD.}
  %
  For each segment $S$ in $\dot S$, let $|S_{nn}|$ be the number of non-null tuples in $S$. The gain $g(\dot S)$ is computed as sum of $g(S)$ weighted by its length $|S_{nn}|$. 
  
  \begin{align}\label{eq:gain}
  g\mbox{(}\dot S\mbox{)} = \sum_{S \in \dot S}
      g(S)\cdot |S_{nn}|
  \end{align}
  the cost $e(\dot S)$ is computed as
  \begin{equation}\label{eq:error}
    e\mbox{(}\dot S\mbox{)} = \max_{S \in \dot S}e\mbox{(}S\mbox{)}
  \end{equation} 
  
  \eat{For each segment $S$ in $\dot S$, let $|S_{nn}|$ be the number of
  non-null tuples in $S$, and $L_S$ be a LMB in $S$. The gain $g\mbox{(}\dot S\mbox{)}$ and the cost $e\mbox{(}\dot S\mbox{)}$ are defined respectively as follows.
  \begin{align}\label{eq:gain}
    \begin{split}
      g\mbox{(}\dot S\mbox{)} = \sum_{S \in \dot S}
      \mbox{(}| t : t &\in L_S,  t \in S|\ -\\[-10pt]
      & |t: t \notin L_S,  t \neq null, t \in S|\mbox{)}\cdot |S_{nn}|
    \end{split}
  \end{align}

  \begin{equation}\label{eq:error}
    e\mbox{(}\dot S\mbox{)} = \max_{S \in \dot S}e\mbox{(}S\mbox{)}
  \end{equation}   
  $e\mbox{(}S\mbox{)}$ is the maximum number of contiguous
  outliers that violate a band OD $\textbf{X} \mapsto_{\Delta} \overline{\textbf{Y}}$ in $S$.
  \begin{equation}~\label{eq:error_in_series}
    \begin{matrix}
      e\mbox{(}S\mbox{)} = \displaystyle \max_{k \in \{ i, \cdots, j \}, 1 \leq i \leq j \leq |S|}  |j
      -i| \textrm{ : }\ t_k \notin L_S, t_k \in S
    \end{matrix}
  \end{equation}}
  \rbox 
  }
  \end{definition}
  
We call band ODs that hold conditionally over subsets of the data and approximately with some exceptions \emph{approximate band conditional ODs} (abcODs). In Equation~\ref{eq:gain}, a gain function rewards correct tuples weighted by the length of series $|S_{nn}|$ excluding tuples with null values
to achieve high recall. Otherwise small series would be ranked high with an extreme optimal case of all segments being individual tuples, which is obviously not desirable.
  
  \begin{example}
  Consider a band OD $\sf{cat\#} \mapsto_{\Delta=1} {\sf{year}} \upA$ and an error rate $\epsilon =1$. Figure~\ref{fig:abcOD_with_piece}(c) visualizes two series based on sequence $T = \{t_1-t_9, t_{15}-t_{22}\}$ in Table~\ref{tab:order} with 
  \eat{$\sf{cat\#} \mapsto_{\Delta=1} \sf{year}\upA{}$ wrt} $S_1=\{t_1 - t_{9}\}$, where $t_2$ is an outlier,
  \eat{$\sf{cat\#} \mapsto_{\Delta=1} \sf{year}\downA{}$ wrt $S_2=\{t_{10}-t_{14}\}$} and \eat{$\sf{cat\#} \mapsto_{\Delta=1} \sf{year}\upA{}$ wrt} $S_2=\{t_{15}-t_{22}\}$, where $t_{18}$ has a missing year. 
  The segmentation $\dot S=\{S_1, S_2\}$ maximizes the optimization function $g(\dot S) = (8 -1) \cdot 9 + 7 \cdot 7= 112$ under a constraint $\max (e(S_1), e(S_3)) = 1 \leq \epsilon$. \rbox
\end{example}

Our problem of abcODs discovery is not a simple matter of finding splitting points. We study a technically challenging joint
optimization problem motivated by real-life applications of finding splits, monotonic bands and approximation (to account for
outliers), which is not easily obtained by simple visualization. Also, note that Fig.~\ref{fig:real-world} presents only a small sample of the data extracted from the entire dataset to illustrate the intuition. In practice, the number of data series can be thousands over large datasets (see Table~\ref{tab:stis}), thus, data cannot be split easily into a few segments. We argue that an automatic approach to discover abcODs is needed as
formulating constraints manually requires domain expertise, is prone to human errors, and is excessively
time consuming.
Automatically discovered dependencies can be manually validated by domain experts, which is a much easier task than manual specification. 
(In our experimental evaluation in Sec.~\ref{sec:experimental_evaluation} it turned out that all the discovered series are true.) 
The purpose of our framework is to alleviate the cognitive burden of human specification.


\subsection{Computing abcODs}
\label{sub:merging_pieces_into_series}
A brute-force solution to the abcOD discovery problem is to consider all possible $2^{n-1}$ segmentations of $n$ tuples over the dataset and select the optimal one as specified in Definition~\ref{def:series}. This has an exponential time complexity, thus, we provide optimizations to compute abcODs more efficiently (in polynomial time). Next, we show that the solution to the abcOD discovery problem in a sequence of tuples contains optimal solutions in subsequences. Thus, the solution to the problem has an optimal substructure property.


\begin{theorem}\label{them:series_sub_structure}
Let ${\sf OPT}(j)$ denote an optimal solution to abcOD discovery problem in $T[j]$ and $T[i, j]$ denotes the subsequence $T[i, ..., j]$. The optimal solution ${\sf OPT}(j), j \in \{ 1, \cdots, n \}$ in a prefix $T[j]$ contains optimal solutions to the subproblems in prefixes $T[1], T[2], \cdots, T[j-1]$. 
\begin{equation}\label{eq:opt}
\fontsize{9.5}{9.5}
  {\sf OPT}(j) = \left \{
    \begin{array}{ll}
      0 &  j = 0  \\[10pt]
      \max_{i \in \{0, \cdots, j-1 \} \text{\space} \mbox{ and }
         \text{\space} e(T[i+1, j]) < \epsilon}\{
      \\[5pt]\quad{\sf OPT}(i) + g(T[i+1, j]) \} & j > 0
    \end{array} \right.
\end{equation}	\rbox  
\end{theorem}

\eatproofs{
\begin{proof}
For subsequence $T[i]$, $i \in [1, j-1]$ with optimal solution $\sf{OPT}(i)$, if we force $T[i+1, j]$ to form a single series with gain $g(T[i+1, j])$, among all segmentations, where $T[i+1, j]$ forms a single series in $T[j]$, there does not exist any segmentations with greater gain than $\sf{OPT}(i)+g(T[i+1,j])$. 
\end{proof}
}

  \begin{algorithm}[htbp]
  \caption{Computing Series} \label{alg:series}
  \SetKw{AAnd}{and}
  \SetKw{New}{new}
  \SetKwFunction{Add}{add}
  \SetKwFunction{LMB}{lmb}
  \SetKwFunction{MaxError}{max\_error}
  \SetKwFunction{SubSeq}{segmnt}
  \SetKwFunction{ComputeLMB}{ComputeLMB}
  \KwData  {$T=\{t_1, t_2, \cdots, t_n\}$, $\Delta$, $\epsilon$}
  \KwResult{segmentation $\dot S$ in $T$}
  $X, G \leftarrow$ two arrays of size $n+1$; $\dot S \leftarrow \emptyset$\;
  $X \leftarrow$ $\emptyset$; $ G \leftarrow \{0, \cdots, 0\}$\;
  \BlankLine
  \For{$j \leftarrow \mbox{\emph{[}}1, n\mbox{\emph{]}}$} {
    \For{ $i \leftarrow \mbox{\emph{[}}1, j-1\mbox{\emph{]}}$} {
      $L_{i+1,j} \leftarrow \ComputeLMB(T[i+1, j], \Delta)$ \; \label{line:alg2_5}
                          $e_{i+1,j} \leftarrow$ cost $e\mbox{(}T\mbox{[}i+1, j\mbox{]}\mbox{)}$;
      $g_{i+1, j} \leftarrow$ gain $g(T[i+1, j])$
      \;\label{line:alg2_6}
      \If{$e_{i+1,j} \leq \epsilon$ \mbox{ and } $\mbox{\emph{(}}G\mbox{\emph{[}}i\mbox{\emph{]}}+g_{i+1,j}) > G\mbox{\emph{[}}j\mbox{\emph{]}}$} {
        $G\mbox{[}j\mbox{]} \leftarrow G\mbox{[}i\mbox{]}+g_{i+1, j}$; $X\mbox{[}j\mbox{]} \leftarrow i$\;\label{line:alg2_8}
      }
    }
  }
	\For{$j \leftarrow n$ \textbf{to} $1$\label{line:alg_series_17}}
	{
	    $i \leftarrow$ $X[j]$; add $S=\{t_i-t_j\}$ into $\dot S$\;
	        $j \leftarrow i-1$ \label{line:alg_series_20}\;
	}
  \Return $\dot S$ \;
\end{algorithm}

\begin{table}[htbp]\centering
\paddingT
\paddingT
  \caption{\small{Computing abcOD.}}
\label{tab:opt}
  \scalebox{0.73}{
      \begin{tabular}{ |c | c | c | c | c | c| c| c| c| c| c| c| c| c| c| c|}
      \hline
      & 1  & 2 & 3  & 4   & 5   & 6   & 7 & 8 & 9 & 15   & 16& 17& ...& 22 \\
      \hline
      \hline
      $T$& '92 &'12  & '96  & '95 & '99 & '00 & '99   &  '01   & '02 &  '82 &'87  &  '89 &... & '92\\
      $G$& 1 & 4  & 5 & 10& 15   & 24   & 35  & 48  & 63 & 64  & 67 & 72 &...&112\\
      $X$& 1 & 1  & 2 & 2&  1  & 1   & 1  & 1 & 1 & 15  & 15& 15&...&15\\
      \hline
    \end{tabular}
  }
\end{table}

For each $i \in [1, j-1]$, a candidate solution with gain $\sf{OPT}(i)+g(T[i+1,j])$ can be found; thus, the optimal solution in a prefix $T\mbox{[}j\mbox{]}$ can be selected among $j$ instead of $2^j-1$ options.
We develop a dynamic-programming
algorithm (Algorithm~\ref{alg:series}) to solve the abcOD discovery problem.
Two arrays are maintained 
of size $n$: array $G$ stores the overall gains of optimal
solutions to the subproblems, i.e.,
$G\mbox{[}j\mbox{]}= {\sf OPT}\mbox{(}j\mbox{)}, j \in \{1, \cdots, n \}$; and array $X$ stores the
corresponding series, i.e., $X\mbox{[}j\mbox{]}$ stores a segment {\sf ID} $i$ that tuples $\{ t_{i}$--$t_{j} \}$ belong to
in a prefix $T\mbox{[}j\mbox{]}, i \in [1, j]$. For each $i \in [1, j-1]$, we consider $T[i+1, j]$ as a single series, compute its gain $g(T[i+1,j])$ and cost $e(T[i+1, j])$ by its LMB $L_{i+1, j}$ (Line~\ref{line:alg2_5}--Line~\ref{line:alg2_6}). If $g(T[i+1,j]) + g(T[i])$ is greater than existing $\sf{OPT}(j)$ and its cost is less than the threshold $\epsilon$, we update $\sf{OPT}(j)$ by $\sf{OPT}(i) + g(T[i+1], j)$, and update segment {\sf ID} $i$ in $X[j]$ (Line~\ref{line:alg2_8}).

The optimal segmentation with {\sf ID}s is collected by scanning segment {\sf ID}s stored in $X$ in reverse order. Let $i = X[n]$, we split $T$ into subsequence $T[1, i-1]$ and $T[i, n]$, and tuples $t_i-t_n$ forms a series. Then, subsequence $T[1, i-1]$ is split in similar fashion until all tuples in $T$ are assigned to a segment (Line~\ref{line:alg1_17}--Line~\ref{line:alg1_20}).

\eat{$T[j]$ can be split into two sub-sequences: $T[i]$ and $T[i+1,j]$, where series in $T[i]$, with gain $g(T[i])$, is computed by {\sf OPT}(i). 
\eat{(1) a
singleton series consisting of $t_j$, and the series in a prefix
$T\mbox{[}j-1\mbox{]}$; (2) a series of length $2$ consisting of $t_j, t_{j-1}$, and the series in a prefix $T\mbox{[}j-2\mbox{]}$; $\cdots$; and finally, a series of length $j$ consisting of all elements in a prefix $T\mbox{[}j\mbox{]}$.}}

\begin{example}\label{ex:opt}
Consider abcOD discovery problem over $T = \{t_1-t_9, t_{15}$--$t_{22}\}$ in Table~\ref{tab:order} given a band OD $\sf{cat\#} \mapsto_{\Delta=1} \sf{year} \upA$ and an error rate $\epsilon =1$. 
\eat{We solve the problem by discovering abcOD in sub-sequences $T[1]$ till $T[22]$. } 
First, $T[1]=\{t_1\}$ is examined. It forms a singleton series with the
  gain ${\sf OPT(1)} =1$ ($G[1]=1$ (as ${\sf OPT}(0)=0$ and  $g(T[1]) = 1$), $X[1]=1$). Next, subsequence $T[2]=\{t_1, t_2\}$ is considered. Tuple $t_2$ can either
  form its own series with the gain equal to $1$ (with the overall gain $1+{\sf OPT}(1)=2$) or be merged into the same series with $t_1$ with the gain $2^2=4$. 
  Thus, $t_2$ and $t_1$ are merged as well as  
  $G[2]=4$ and $X[2]=1$. Wrt tuple $t_3$, there are three candidates to the optimal solution in $T[3]$: (1) $t_3$ forms a single series ($g(t_3)=1$) and optimal solution $\{t_1, t_2\}$ in $T_2$ remains ($g(T[2])=4$), thus, the total gain is $1+4$; (2) $\{t_2, t_3\}$ forms a series ($g(t_{2-3})=2^2=4$) and optimal solution $\{t_1\}$ in $T_1$ remains ($g(T[1])=1$), in which case the total gain is also $4+1=5$; and (3) $\{t_1-t_3\}$ forms a series, where $t_2$ (or $t_1$ / $t_3$) is an outlier, hence, the total gain is $(2-1)\times 3 = 3$. Among the above candidates  option (1) (or (2)) is chosen with maximal gain, and $G[3]=5$, $X[3]=2$ (or $X[3]=1$). 
 The rest of tuples are processed accordingly with the results reported in Table~\ref{tab:opt}.
 To output all series, the array $X$ is checked in reverse order. Given that $X[22]=15$, an optimal solution is achieved by a series consisting of $t_{15}-t_{22}$ and an optimal solution in $T[9]$; given $X[9]=1$, optimal solution in $T[9]$ is achieved by a series consisting also of $\{t_{1} - t_{9}\}$. 
 \rbox
\end{example}



\begin{theorem}\label{theo:series}
  Algorithm~\ref{alg:series} solves the abcOD discovery problem optimally in $O\mbox{(}n^3\log n\mbox{)}$ time in a sequence $T$ of size $n$. \rbox
\end{theorem}

\eatproofs{
\begin{proof}
Algorithm~\ref{alg:series} applies dynamic programming to solve Eq/-uation~\ref{eq:opt}, which is proved in Theorem~\ref{them:series_sub_structure}. 
The recurrence in Equation~\ref{eq:opt} specifies that the optimal solution in subsequence $T\mbox{[}i\mbox{]}$ are selected among $i$ alternative options: (1) a
singleton series consisting of $t_i$, and the optimal solution in subsequence
$T\mbox{[}i-1\mbox{]}$; (2) a series of length $2$ consisting of $\{ t_i, t_{i-1} \}$, and the optimal solution in subsequence $T\mbox{[}i-2\mbox{]}$, etc.; and finally, a series of length $i$ consisting of all tuples in subsequence $T\mbox{[}i\mbox{]}$. 
It requires $O(n)$ iteration to process subsequence $T[i]$, $i \in [1, n]$, where each iteration takes time $O\mbox{(}n\log n\mbox{)}$ based on Theorem~\ref{lemma:lmb}. Thus, Algorithm~\ref{alg:series} solves the abcOD discovery problem in time $O(n^3\log n)$.
\end{proof}
}

%% file: pieces.tex
\subsection{Pruning with Pieces}
\label{sub:pieces}


To further prune the search space, we develop another optimization for the abcOD discovery solution based on the idea of {\em splitting and segmenting}. A sequence of tuples is split into \emph{pieces}, which are contiguous subsequences of tuples that are monotonic within band-width $\Delta$. Then, abcOD discovery segmentation is computed based on pieces to speed up the performance without sacrificing optimality. To split a sequence $T$ into pieces, we first introduce the notion of {\em pre-pieces}. To distinguish outliers and boundary tuples between adjacent series, 
we consider bidirectional orders in the definition of pre-pieces, since they can potentially overlap even for case of discovery of unidirectional abcODs.

\eat{To further prune the search space, we develop a greedy discovery algorithm  that is based on {\em pieces}. Pieces split a
sequence of tuples based on pre-pieces into contiguous
subsequences that are monotonic within $\Delta$ to speed up the performance without sacrificing the precision
(Sec.~\ref{exp:effect}).} 
{
\begin{definition}[Pre-Piece]\label{def:1}
  Given a sequence $T$ of $n$ tuples and a marked list of
  attributes $\overline{\textbf{Y}}$, a contiguous subsequence
  $T'=\{t_{i}, t_{i+1},..., t_{j}\}$ is a {\em pre-piece} (PP) if (1) $\forall_{k, m \in \{i, \cdots, j \}, k < m }$
  $t_{k} \preceq_{\Delta, \overline{\textbf{Y}}} t_{m} \mbox{ or } t_k.\textbf{Y} = {\sf null}$\eat{$\forall_{i \leq k < j}$
  $t_k \preceq_{\Delta , \overline{\textbf{Y}}} t_{k+1} \mbox{ or }
  t_k = {\sf null} \mbox{ or }$ or $t.\textbf{Y} = {\sf null}$},
  and (2) $T'$ cannot be extended without violating the property
  (1). $T'$ is called an {\em increasing pre-piece} (IP) if
  $\overline{\textbf{Y}} = \textbf{Y}\upA$ and a \emph{decreasing
    pre-piece} (DP) if $\overline{\textbf{Y}} = \textbf{Y}\downA$.
  \rbox
\end{definition}
}

\begin{example}\label{ex:pre-pieces}
  Let band-width $\Delta = 1$ and $\overline{\textbf{Y}} = [\sf{year}] \upA$ or $\overline{\textbf{Y}} = [\sf{year}] \downA$.  Consider a
  sequence of tuples $T = \{t_1-t_9, t_{15}-t_{22}\}$ over
  Table~\ref{tab:order} ordered by an attribute $\sf{cat\#}$. There
  are five pre-pieces in $T$, i.e., $\{t_1$--$t_2\}$,
  $\{t_2$--$t_4\}$, $\{t_3$--$t_{9}\}$, $\{t_{8}, t_{9}, t_{15}\}$ and
  $\{t_{15}$--$t_{22}\}$ as marked with dashed lines in Fig.~\ref{fig:abcOD_with_piece}(a).\rbox
\end{example}

\eat{
\begin{figure}[tbp]\centering
  \includegraphics[scale=0.5]{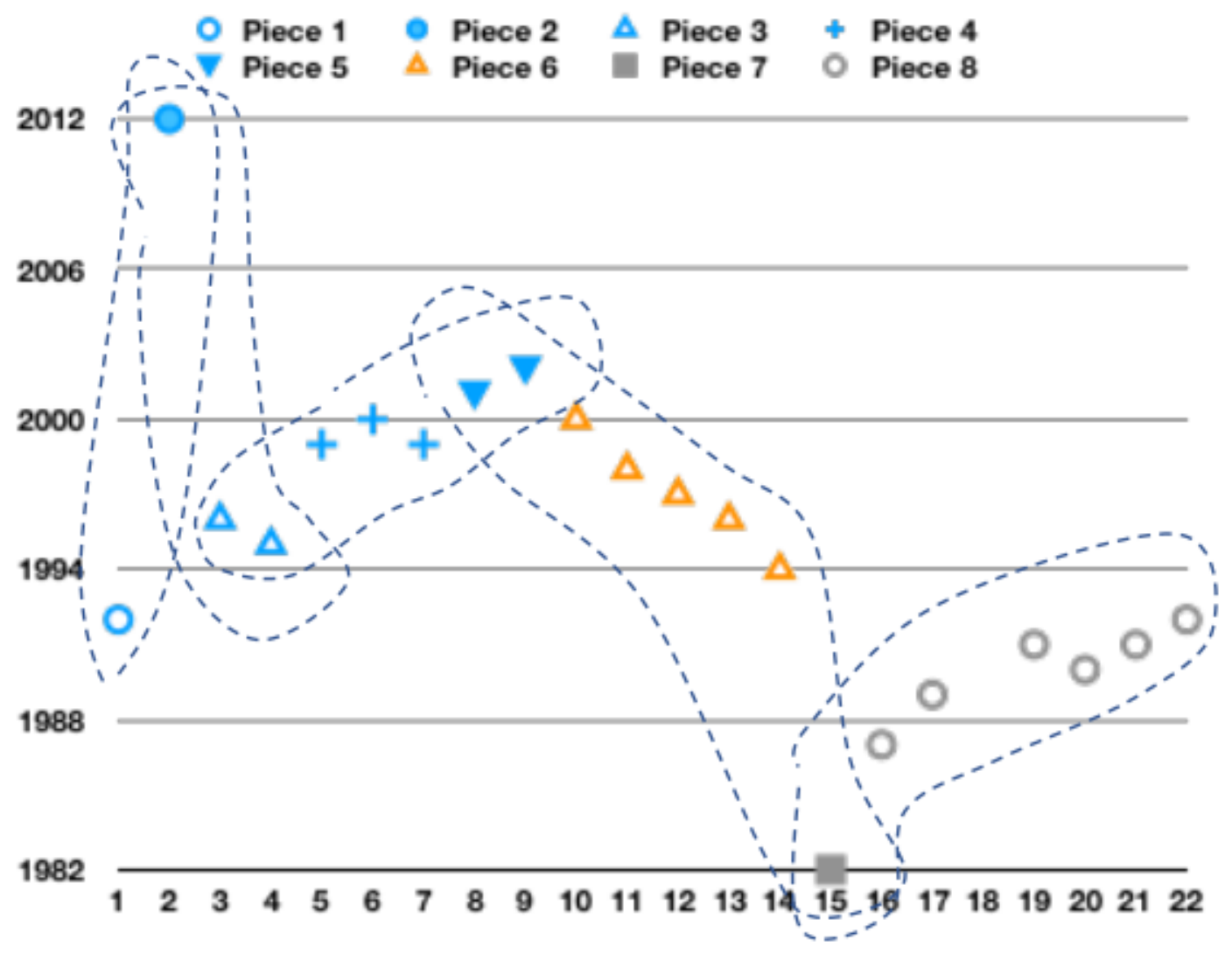}
  \caption{\small{Pieces and pre-pieces \colB{(marked by dash-line)}. 
    \label{fig:piece}}}
\end{figure}
}

A pairwise-disjoint set of \emph{pieces} is obtained by separating the intersections of all pre-pieces. 

\begin{definition}[Piece]
  A \emph{piece} $P$ is a
  subsequence in a sequence of tuples $T$ such that: (1)
  non-overlapping tuples from a pre-piece with other pre-pieces create a separate piece, and 
  (2) overlapping tuples between pre-pieces create a separate piece.\rbox
\end{definition}	

\begin{example}\label{ex:piece} 
Continuing Example~\ref{ex:pre-pieces} 
%
  Fig.~\ref{fig:abcOD_with_piece}(b) 
  illustrates pieces in $T = \{t_1-t_9, t_{15}-t_{22}\}$:
  $P_1=\{t_1\}, P_2=\{t_2\}, P_3=\{t_3, t_4\}$, $P_4=\{t_5$--$t_7\},
  P_5=\{t_{8}, t_{9}\}, P_6=\{t_{15}\}$ and $P_7=\{t_{16}$--$t_{22}\}$. \rbox
\end{example}



The pseudo-code of the algorithm to compute pieces is presented in Algorithm~\ref{alg:piece}. Pre-pieces are constructed by scanning iteratively tuples in a sequence $T$. Tuple $t \in T$ can both extend existing IPs and (or) DPs, and form a new IP and (or) DP. To make sure each PP is contiguous and cannot be extended, maximal lengths of IPs and DPs are stored in two arrays of length $n$: $M_{\sf inc}$, $M_{\sf dec}$, respectively; i.e., among all IPs (DPs) ending at $t_j$, in $M_{\sf inc}[i]$ the length of the maximal IP ({\sf DP}) is stored, if its best tuple is $t_i$ ($M_{\sf dec}[i]$), $i \leq j$. Given that PPs must be contiguous, for tuple $t_j$, the trace of PPs extended by $t_{j-1}$ is kept. Assuming a IP ending at $t_{j-1}$ is a maximal IP, i.e., it cannot be extended by tuples in $T[j-1]$, ff tuple $t_j$ can extend the IP, its length is updated in $M_{\sf inc}[{\sf max}]$, where $t_{\sf max}$ is the new best tuple (Line~\ref{line:alg3_11}); if $t_{\sf max}$ is different from $t_j$, a new IP of length $1$ is kept with best tuple $t_j$ (Line~\ref{line:alg3_8}). Similarly,  the length of DP is encoded with best tuple $t_i$ in $M_{\sf dec}[i]$.
If $t_j$ cannot extend IP with length $l_{\sf inc}$ ending at $t_{j-1}$, a maximal $\sf IP=\{t_{j-l_{inc}} - t_j-1\}$ is found in subsequence $T[j]$. If it is longer than the longest DP so far, a PP is found in $T$ (Line~\ref{line:alg3_13}). Given that pieces are obtained by separating intersections of pre-pieces, a sequence $T$ is split by the first and last tuples of each PP (Line~\ref{line:alg3_18}).

\begin{figure}[t]\centering
  \includegraphics[scale=0.27]{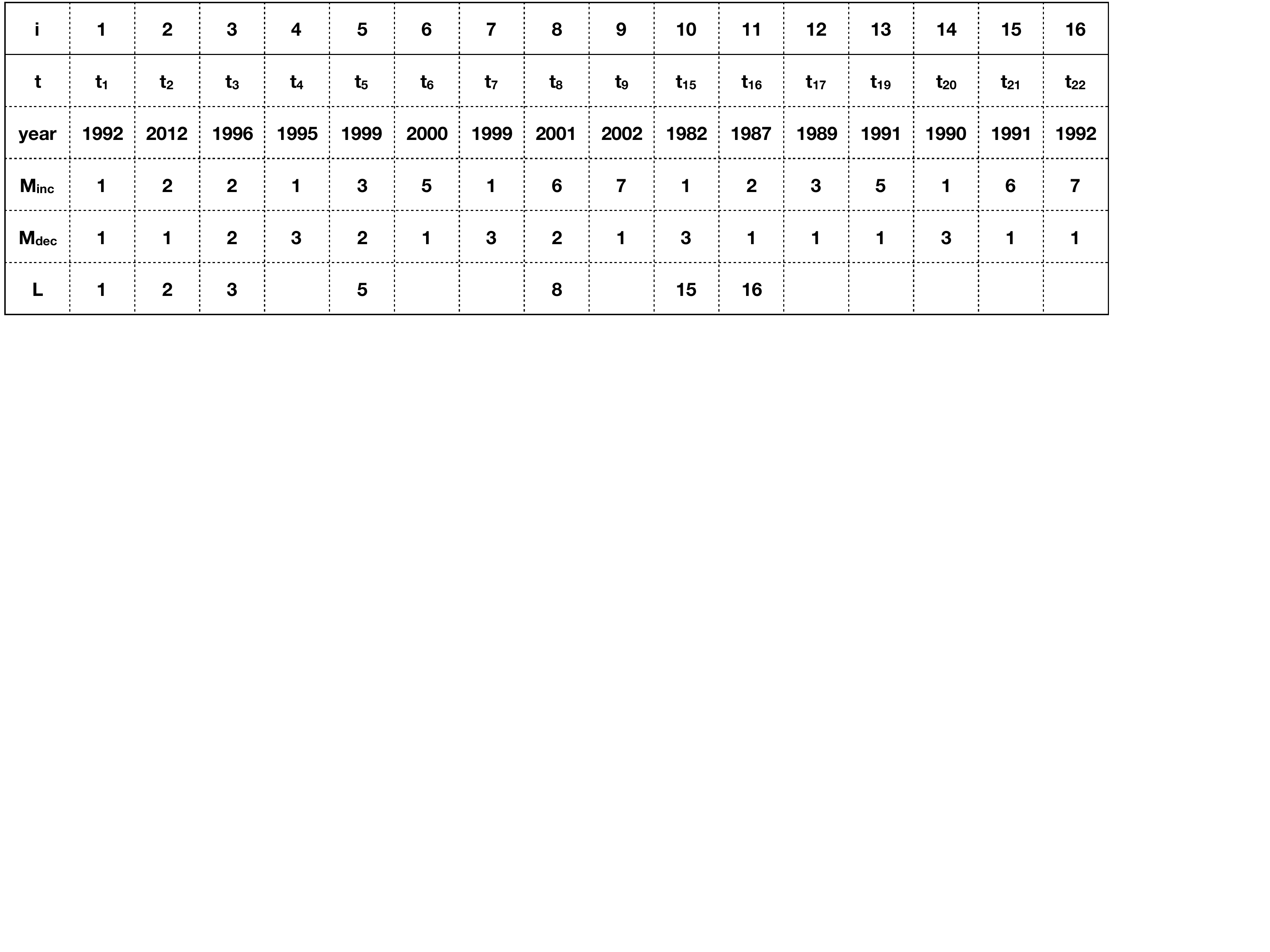}
 \vspace{-5.2cm}
 \paddingT
  \caption{\small{Details for pieces computation.
    \label{fig:piece_computation_example}}}
    \paddingD
\end{figure}

\begin{example}\label{ex:compute_piece}
Following Example~\ref{ex:piece}, when processing $t_1$, an IP and a DP are found both with the maximal tuple $t_1$ and the length of one ($M_{\sf inc}[1] = M_{\sf dec}[1] = 1$). Next, $t_2$ is processed, which
  extends a current IP with the maximal tuple $t_1$, and forms a new DP of length $1$ ($M_{\sf inc}[2] = 2, M_{\sf dec}[2] = 1$). When $t_3$ is processed, it cannot extend existing IP of length $2$ that is longer than existing DP with length $1$; i.e., a {\sf PP} $\{t_1-t_2\}$ is found. $M_{\sf inc}[3]=1$ is updated and the pre-piece boundary indexes $1, 3$ are inserted into array $L$. The rest of the tuples are processed accordingly (Figure~\ref{fig:piece_computation_example}) with pieces boundaries obtained by scanning $L$ reported in Figure~\ref{fig:abcOD_with_piece}(b).
\end{example}

\begin{lemma}\label{lemma:piece}
  Algorithm~\ref{alg:piece} takes $O\mbox{((}\Delta +1\mbox{)}\cdot n\mbox{)}$ time to compute pieces in a sequence $T$ of size $n$. \rbox
\end{lemma}

  \begin{algorithm}[htbp]
  \caption{Compute Pieces} \label{alg:piece}
  \SetKw{Max}{max}
  \SetKw{Min}{min}
  \SetKwFunction{UpdateMax}{updateIfMax}
  \SetKwFunction{UpdateMin}{updateIfMin}
  \SetKwFunction{Remove}{remove}
  \SetKwFunction{SubSeq}{sub\_seq}
  \SetKwFunction{Split}{split}
  \SetKwFunction{ValueSet}{valueSet}
  \SetKwFunction{Append}{append}
  \SetKwFunction{Extend}{extend}
  \SetKwFunction{Add}{add}
  \SetKwFunction{Next}{next}
  \SetKwFunction{Head}{head}
  \SetKwFunction{Insert}{insert}
  \KwData{$T=\{t_1, t_2, \cdots, t_n\}$, $\Delta $} 
  \KwResult{the set of pieces $\dot P$ in $T$}
  $M_{{\sf ins}} \leftarrow \emptyset$; $M_{{\sf dec}} \leftarrow \emptyset$;
  $\dot P \leftarrow \emptyset$; $L \leftarrow$ array of size $n$\;
  $i\leftarrow 0$\;
	\BlankLine

	\For{\textbf{each} $j \leftarrow \mbox{\emph{[}}1, n\mbox{\emph{]}}$}
	{
	    $I_{\sf inc} \leftarrow$ indices in $M_{\sf inc}$ updated by $j-1$\;
	    $I_{\sf dec} \leftarrow$ indices in $M_{\sf dec}$ updated by $j-1$\;
	    $l_{\sf inc} \leftarrow$ maximal value in $M_{\sf inc}$ updated by $j-1$\;
	        $l_{\sf dec} \leftarrow$ maximal value in $M_{\sf dec}$ updated by $j-1$\;
	    \For{\textbf{each} $i \in I_{\sf inc}$}
	    {
	       $M_{\sf inc}[j] \leftarrow \max(M_{\sf inc}[j], 1)$\; \label{line:alg3_8}
	       \uIf{$t_i \preceq_{\Delta, \overline{\textbf{Y}}} t_j$}
	       {
	        $t_{\sf max} \leftarrow \max_{\textbf{Y}}(t_i, t_j)$\;
	        update $M_{\sf inc}[\sf max]$ by $\max(M_{\sf inc}[\sf max], M_{\sf inc}[i]+1)$\; \label{line:alg3_11}
	       }{
	        \uElseIf{$l_{\sf inc} > l_{\sf dec}$}{
	        insert $j, j-l_{\sf inc}$ into $L$\; \label{line:alg3_13}
	        }
	       }
	    }
	    \For{\textbf{each} $i \in I_{\sf dec}$}
	    {
	       $M_{\sf dec}[j] \leftarrow \max(M_{\sf dec}[j], 1)$\;
	       \uIf{$t_i \preceq_{\Delta, \overline{\textbf{Y}}} t_j$}
	       {
	        $t_{\sf min} \leftarrow \min_{\textbf{Y}}(t_i, t_j)$\;
	        update $M_{\sf dec}[\sf min]$ by $\max(M_{\sf dec}[\sf min], M_{\sf dec}[i]+1)$\; 
	       }{
	        \uElseIf{$l_{\sf dec} > l_{\sf inc}$}{
	        insert $j, j-l_{\sf dec}$ into $L$\;
	        }
	       }
	    }
	  }

\While{$i \leq n$}
{
    $\dot P.\Add(\{t_{L[i]} - t_{L[i+1]}\})$\;\label{line:alg3_18}
	$i \leftarrow i+1$\;
}

	\Return $\dot P$ \;
\end{algorithm}

\begin{table}[t]\centering
	\vspace{-0.2in}
  \caption{{\small{Computing abcODs with Pieces.}}}
  \label{tab:greedy}
  \scalebox{0.8}{
    \begin{tabular}{l|l|l|l|l|l|l|l}
      \hline
      & $g\mbox{(}P_{0}\mbox{)}$ & $g\mbox{(}P_{1}\mbox{)}$ & $g\mbox{(}P_{1-2}\mbox{)}$ & $g\mbox{(}P_{1-3}\mbox{)}$ & $g\mbox{(}P_{1-4}\mbox{)}$ & $g\mbox{(}P_{1-5}\mbox{)}$ & $g\mbox{(}P_{1-6}\mbox{)}$ \\
      \hline
      \hline
      $g\mbox{(}P_{1}\mbox{)}$ & $\bf 1$ &&&&&&\\
      $g\mbox{(}P_{1-2}\mbox{)}$ & $\bf 4$ & $2$ &&&&&\\
      $g\mbox{(}P_{1-3}\mbox{)}$ & $ 8$ & $\bf 10$ & $ 8$ &&&&\\
      $g\mbox{(}P_{1-4}\mbox{)}$ & $\bf 35$ & $25$ & $29$ & $19$ &&&\\
      $g\mbox{(}P_{1-5}\mbox{)}$ & $\bf 63$ & $49$ & $53$ & $35$ & $39$&&\\
      $g\mbox{(}P_{1-6}\mbox{)}$ & $60$ & $46$ & $52$ & $34$ & $44$& $\bf 64$&\\
      $g\mbox{(}P_{1-7}\mbox{)}$ & $-$ & $-$ & $-$ & $-$ & $-$& $\bf 112$& $100$\\
      \hline
    \end{tabular}
  }
\end{table}

Pieces are used to prune the search space for the abcODs discovery (Algorithm~\ref{alg:series_with_pieces}). Instead of processing each tuple individually, the sequence is processed piece by piece (Line 6--Line 9).
Alg.~\ref{alg:series_with_pieces} extends Alg.~\ref{alg:series} by additionally keeping tuples in the same piece always within the same segment (Line 11). Note that multiple pieces can belong to the same series. 


\begin{example}\label{ex:greedy}
Consider abcOD discovery problem over $T = \{t_1-t_9, t_{15}$--$t_{22}\}$ in Table~\ref{tab:order} given $\sf{cat\#} \mapsto_{\Delta=1} \sf{year} \upA$ and an error rate $\epsilon =1$.
%
Fig.~\ref{fig:abcOD_with_piece}(b) illustrates pieces $P_1$--$P_7$ (Example~\ref{ex:compute_piece}). A piece $P_1=\{t_1\}$ examined first forms a singleton series with the
  gain $g(P_1) =1$. 
  Piece $P_2$ can either form its own series with overall gain equal to $2$, or can be merged with $P_1$ with gain equal to $4$, which is the optimal solution in subsequence $P_{1-2}$.
 The rest of pieces are processed accordingly with results reported in Table~\ref{tab:greedy}, where the optimal solution is highlighted in each subsequence. Figure~\ref{fig:abcOD_with_piece}(c) illustrates the series over pieces $\{P_1$--$P_7\}$. \rbox
\end{example}


In practice, pieces are large, hence, the number of pieces is small (i.e., $m \mbox{ }\ll \mbox{ } n$), which leads to an efficient and optimal algorithm for abcOD discovery.

\begin{theorem}\label{theo2:greedy}
Algorithm~\ref{alg:series_with_pieces} finds optimal solution for abcODs discovery problem in $O\mbox{(}m^2 n \log n\mbox{)}$
  time, where $m$ is the number of pieces in $T$, and $n$ is the
  number of tuples in $T$. \rbox
\end{theorem}

\eatproofs{
\begin{proof}
Assume $T[i]$ ends at piece $P_i=\{t_{i-m+1}, \cdots, t_i\}$ of length $m$. Every tuple in $P_i$ belongs to
  the same sets of pre-pieces, there are no outliers that violate a 
  LMB
  in $P_i$, i.e., $g\mbox{(}T\mbox{[}i-m+1, i]\mbox{)} = m^2$\eat{ and
  $g\mbox{(}T\mbox{[}i-k+1, i\mbox{]}\mbox{)} = k^2, k \in [1, m]$}.
If the algorithm does not find the optimal solution, then there
  exists a tuple $t_{i-k} \in P_i, 1 \leq k \leq m-1$ that splits
  $P_i$ into two series: $\{t_{i-m+1}, \cdots, t_{i-k}\}$ and
  $\{t_{i-k+1}, \cdots, t_i\}$, where the profit is
  ${\sf OPT}\mbox{(}i-k\mbox{)} + k^2$.  By contradiction this
  assumption does not hold, i.e.,
  ${\sf OPT}\mbox{(}i\mbox{)} - {\sf OPT}\mbox{(}i-k\mbox{)} \geq k^2
  $.
Let tuple $t_{i-j+1}$ be the first tuple in the last series
  $S_{i-m}$ of the optimal solution ${\sf OPT}\mbox{(}i-m\mbox{)}$,
  where the length of a LMB
  in series $S_{i-m}$ is $l$, i.e.,
  $j \geq m + 1, l >0$; and the maximal number of consecutive outliers
  in $T\mbox{[}i-m\mbox{]}$ is $q$. According to
  Theorem~\ref{them:best}, $\{t_{i-m+1}, \cdots, t_{i}\}$ extends the
  length of 
  LMB
  in $S_{i-m}$ by $m$ without increasing $q$. That is,
${\sf OPT}\mbox{(}i\mbox{)} = {\sf OPT}\mbox{(}i-j\mbox{)} + \mbox{(}l+m\mbox{)}^2$.
Similarly,
${\sf OPT}\mbox{(}i-k\mbox{)} = {\sf OPT}\mbox{(}i-j\mbox{)} + \mbox{(}l+m-k\mbox{)}^2$.
This means that 
${\sf OPT}\mbox{(}i\mbox{)} - {\sf OPT}\mbox{(}i-k\mbox{)} =
\mbox{(}l+m\mbox{)}^2 - \mbox{(}l+m-k\mbox{)}^2 =
2k\mbox{(}l+m\mbox{)} > k^2$, which leads to the contradiction.

Algorithm~\ref{alg:series_with_pieces} first finds all pieces in the sequence $T$ of length $n$, which takes time $O\mbox{(}\Delta + 1\mbox{)}\cdot n$. Assume the number of pieces is $m$, the algorithm applies dynamic programming on $m$ pieces, similarly as Algorithm~\ref{alg:series}, which takes time $O\mbox{(}m^2n\log n\mbox{)}$, as computing LMBs takes time $O(n \log n)$. Therefore, the overall time complexity is $O\mbox{(}m^2n\log n\mbox{)}$.
\end{proof}
}




  \begin{algorithm}[t]
  \caption{Computing Series with Pieces} \label{alg:series_with_pieces}
  \SetKw{AAnd}{and}
  \SetKw{New}{new}
  \SetKwFunction{Add}{add}
  \SetKwFunction{LMB}{lmb}
  \SetKwFunction{MaxError}{max\_error}
  \SetKwFunction{SubSeq}{segmnt}
  \SetKwFunction{ComputePieces}{ComputePieces}
  \SetKwFunction{ComputeLMB}{ComputeLMB}
  \KwData  {$T=\{t_1, t_2, \cdots, t_n\}$, $\Delta$, $\epsilon$}
  \KwResult{segmentation $\dot S$ in $T$}
  $\dot P \leftarrow \{P_1, P_2, \cdots, P_m\}$ by $\ComputePieces(T, \Delta)$\;  
  $X, G \leftarrow$ two arrays of size $m+1$; $\dot S \leftarrow \emptyset$\;
  $X \leftarrow$ $\emptyset$; $ G \leftarrow \{0, \cdots, 0\}$\;
  \BlankLine
  \For{$j \leftarrow \mbox{\emph{[}}1, m\mbox{\emph{]}}$} {
    \For{ $i \leftarrow \mbox{\emph{[}}1, j-1\mbox{\emph{]}}$} {
    $L_{i+1,j} \leftarrow \ComputeLMB(P[i+1, j], \Delta)$ \; \label{line:alg5_5}
                          $e_{i+1,j} \leftarrow$ cost $e\mbox{(}P\mbox{[}i+1, j\mbox{]}\mbox{)}$;
      $g_{i+1, j} \leftarrow$ gain $g(P[i+1, j])$
      \;\label{line:alg5_6}
      \If{$e_{i+1,j} \leq \epsilon$ \mbox{ and } $\mbox{\emph{(}}G\mbox{\emph{[}}i\mbox{\emph{]}}+g_{i+1,j}) > G\mbox{\emph{[}}j\mbox{\emph{]}}$} {
        $G\mbox{[}j\mbox{]} \leftarrow G\mbox{[}i\mbox{]}+g_{i+1, j}$; $X\mbox{[}j\mbox{]} \leftarrow i$\;\label{line:alg2_8}
      }
    }
  }
	\For{$j \leftarrow n$ \textbf{to} $1$\label{line:alg_series_17}}
	{
	    $i \leftarrow$ $X[j]$; add $S=\{P_i-P_j\}$ into $\dot S$\;
	        $j \leftarrow i-1$ \label{line:alg_series_20}\;
	}
  \Return $\dot S$ \;
\end{algorithm}

%% file: bidirectional.tex
\section{Bidirectional 
Discovery} \label{sec:bidDiscovery}
Unidirectional abcODs are most common in practice (as we verified experimentally in Sec.~\ref{sec:experimental_evaluation}), however, in some applications one needs the additional semantics of \emph{ascending} and \emph{descending} orders. Thus, for generality we extend our formalism to bidirectional abcODs discovery. To deal with bidirectionality, we introduce the concepts of (longest) increasing and decreasing bands (following Definition~\ref{def:2}).


\begin{definition}[increasing and decreasing bands]\label{def:bi_lmb}
  A monotonic band is called an \emph{increasing band} (IB) if   $\overline{\textbf{Y}} = \textbf{Y}\upA$ (and
  a longest IB (LIB) if it is the longest amongst IBs)  and a decreasing
    band (DB) if
  $\overline{\textbf{Y}} = \textbf{Y}\downA$ (and a \emph{longest DB} (LDB) if it is the longest amongst DBs). \rbox
\label{def:lmb_bidirectional}
\end{definition}

\begin{example}\label{ex:series}
  Consider a band OD ${\sf cat\#}$ $\mapsto_{\Delta = 1}$ $\overline{{\sf year}}$ over  Table~\ref{tab:order} ordered by ${\sf cat\#}$. Suppose tuples
  $T$ = $\{ t_{10}$--$t_{14} \}$ form one series. 
  There is a LDB $\{ t_{10}$--$t_{14} \}$ (with {\sf year}
  $\{'00, '98, '97, '96, '94\}$)  in $T$ and there are two LIBs $\{ t_{11}, t_{12} \}$ ($t_{11} \preceq_{\Delta=1,\sf{year}\upA{}} t_{12}$ holds with {\sf year} $\{'98, '97\}$)  and $\{ t_{12}, t_{13} \}$ ($t_{12} \preceq_{\Delta=1,\sf{year}\upA{}} t_{13}$ holds with {\sf year} $\{'97, '96\}$) in $T$.  Thus, 
  a LMB over $T$ consisting of tuples $\{ t_{10}$--$t_{14} \}$ forms a LDB. 
%
  \eat{Based on Definition~\ref{def:lmbSeq}, 
since 
    $t_{10} \preceq_{\Delta=1, \sf{year}\upA{}} t_{11}$,
    $t_{11} \preceq_{\Delta=1,\sf{year}\upA{}} t_{13}$ and 
    $t_{11} \preceq_{\Delta=1, \sf{year}\upA{}} t_{14}$ 
do not hold, tuples $t_{10}$, $t_{13}$ and $t_{14}$ are not part of the LIB with tuples $t_{11}$ and $t_{12}$. However,  
    $t_{10} \preceq_{\Delta=1,\sf{year}\downA{}} t_{11}$,
    $t_{10} \preceq_{\Delta=1,\sf{year}\downA{}} t_{12}$,
    $t_{10} \preceq_{\Delta=1,\sf{year}\downA{}} t_{13}$,
    $t_{10} \preceq_{\Delta=1,\sf{year}\downA{}} t_{14}$,
    $t_{11} \preceq_{\Delta=1,\sf{year}\downA{}} t_{12}$, 
    $t_{11} \preceq_{\Delta=1,\sf{year}\downA{}} t_{13}$,
    $t_{11} \preceq_{\Delta=1,\sf{year}\downA{}} t_{14}$,
    $t_{12} \preceq_{\Delta=1,\sf{year}\downA{}} t_{13}$,
    $t_{12} \preceq_{\Delta=1,\sf{year}\downA{}} t_{14}$,
    $t_{13} \preceq_{\Delta=1,\sf{year}\downA{}} t_{14}$
    .} 
  \rbox
\end{example}

To compute a LMB both a LIB (Section~\ref{sec:longest_increasing_band})
and a LDB need to be computed, and the one with longer length is chosen as a LMB. Computing a LDB in $T$ is symmetrical as calculating a LIB, where the best tuple of DBs of length $k+1$ in $T[i+1]$, denoted as $l_{k+1,i+1}$ satisfies the following recurrence, $v = \max_\textbf{Y} \mbox{(}l_{k+1, i}$,
  $\min_\textbf{Y}(t_{i+1}$,
  $l_{k, i}))$.

\begin{align*}\label{eq:best_tuple_ldb}\fontsize{9.5}{9.5}
    \begin{split}
      & l_{k+1, i+1} =
       \left \{
        \begin{array}{rl}
          v &  l_{k, i} \preceq_{\Delta, \textbf{Y} \downA} t_{i+1}\\
        l_{k+1,i} & \emph{otherwise}
        \end{array}
      \right.		
    \end{split}
  \end{align*}	

The adapted Algorithm~\ref{alg:series} to consider symmetry between IBs and DBs remains optimal for discovery of bidirectional abcODs.


\begin{theorem}
\label{thm:optBid}
Extended Algorithm~\ref{alg:series}  solves bidirectional abcOD discovery problem optimally in $O\mbox{(}n^3\log n\mbox{)}$ time over a sequence of tuple $T$ of size $n$.
\rbox
\end{theorem}

However, pieces-based algorithm (Algorithm~\ref{alg:series_with_pieces}) may produce sub-optimal solutions over datasets with bidirectional abcODs, when adjacent increasing and decreasing pre-pieces are near symmetric with erroneous values on the borders. 

\begin{figure}[t]\centering
  \includegraphics[scale=0.45]{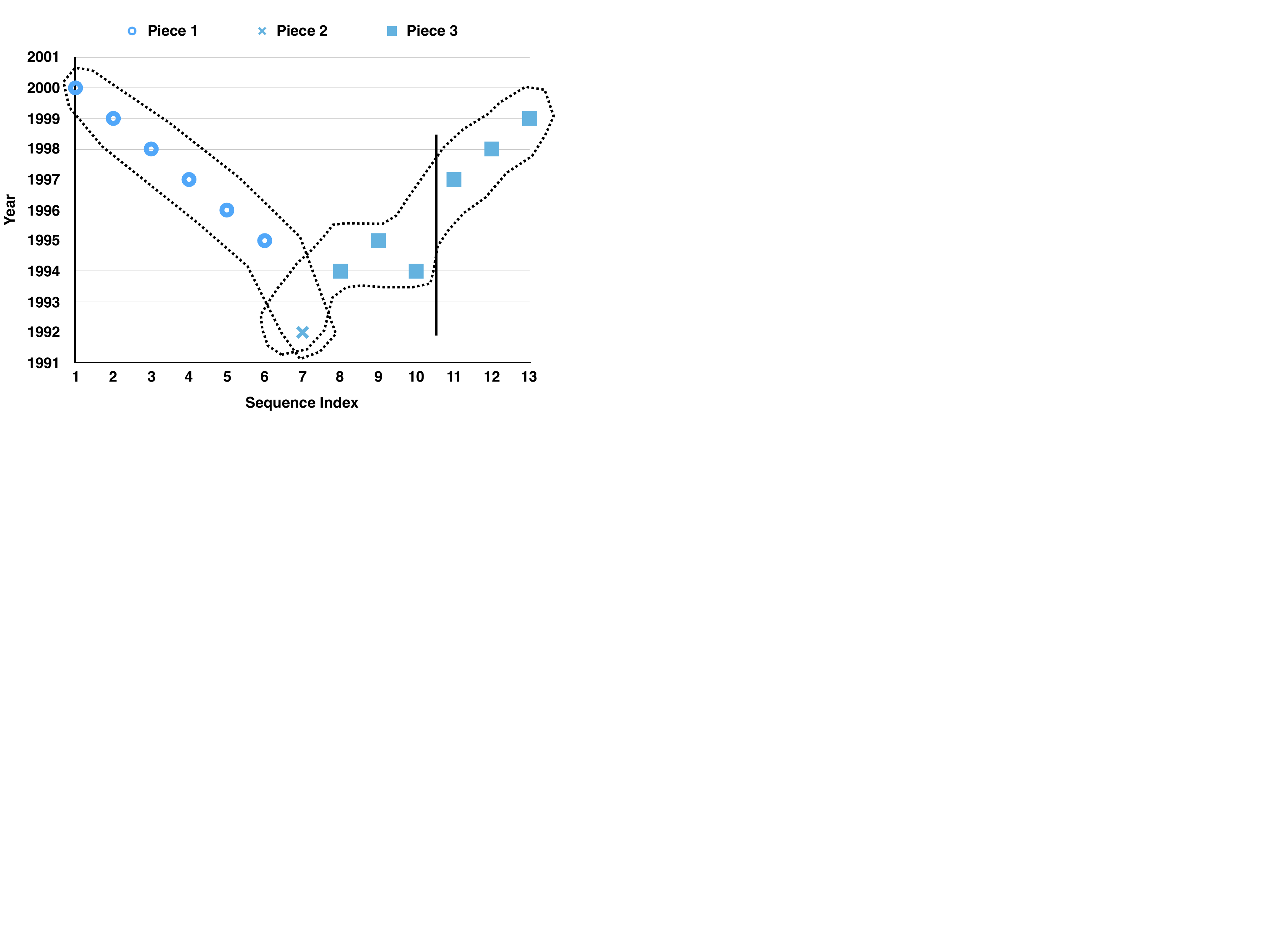}
\vspace{-2.8in}
    \paddingT
  \caption{\small{Counter example. 
   \label{fig:pieces_counter_example}}}
   \paddingD
\end{figure}

\begin{example}\label{eg:sub-optimal}
Consider a sequence of tuples $T=\{t_1$--$t_{13}\}$ in Fig.~\ref{fig:pieces_counter_example}, where attribute ${\sf cat\#}$ corresponds to sequence index over a bidirectional abcOD $\sf{cat\#} \mapsto_{\Delta} \overline{[{\sf year}]}$. Let $\Delta = 1$ and $\epsilon = 1$. As shown in  Fig.~\ref{fig:pieces_counter_example}, there are two pre-pieces in $T$ (denoted with dash lines), and thus, three pieces: $P_1=\{t_1$--$t_6\}, P_2=\{t_{7}\}, P_3=\{t_{8}$--$t_{13}\}$.
Alg.~\ref{alg:series_with_pieces} finds two non-optimal solutions in $T$ with the same gain of $85$: 1) $S_1=\{t_1$--$t_7\}$ and $S_2=\{t_8$--$t_{13}\}$; and 2) $S_1=\{t_1$--$t_6\}$ and $S_2=\{t_7$--$t_{13}\}$. But the solution by Alg.~\ref{alg:series} proven to be optimal for discovery of bidirectional abcODs (Theorem~\ref{thm:optBid}) is: $S_1=\{t_1$--$t_{10}\}$ and $S_2=\{t_{11}$--$t_{13}\}$ with a higher gain equal to $89$. \rbox
\end{example}

The above case is rare in real-world applications. Experimental results in Sec.~\ref{exp:effect} illustrate that the pieces-based algorithm does not sacrifice precision in practice over bidirectional abcODs. 

%% file: exp.tex
\section{Experimental Evaluation} \label{sec:experimental_evaluation}

\subsection{Data Characteristics and Settings} 
\label{sub:experiment_settings}

\smallskip 
\noindent{\textbf{Datasets.}} We use two real-world datasets for experiments (1) \emph{Music} (Footnote 1), and (2) \emph{Car} (Footnote 2). The collected \emph{Music} dataset has 1M tuples. It contains information about music releases over 100 years 
including attributes ${\sf label}$, $\sf{title}$, ${\sf
  country}$, ${\sf artists}$, $\sf{genres}$, ${\sf cat\#}$, ${\sf format}$, ${\sf
  year}$ and ${\sf month}$. The \emph{Car} dataset has 362
tuples. It contains information about second-hand cars including attributes ${\sf year}$, $\sf{vehicle \mbox{ } identification \mbox{ } number}$ (${\sf VIN}$), $\sf{order id}$, $\sf{description}$, $\sf{model}$, $\sf{link}$, $\sf{block id}$ and $\sf{car type}$. 
Whenever it is not stated otherwise, we report the results with respect to (bidirectional) abcODs ${\sf cat\#}$ $\mapsto_{\Delta}$ $\overline{{\sf year}}$ over the \emph{Music} dataset and ${\sf VIN}$ $\mapsto_{\Delta}$ $\overline{{\sf year}}$ over \emph{Car} dataset that we automatically identified as candidates for embedded band ODs (as described in Section~\ref{sub:data_segmentation_into_series}).


\smallskip
\noindent{\underline{Real-world Datasets.}} We categorize the real-world
data into five groups by sampling the datasets. Table~\ref{tab:stis} shows statistics for the sampled datasets with real-world errors; SS denotes series size, MV missing values and IV incorrect values. Inc denotes percentage of tuples in unidirectional (ascending) series. Note that over the entire \emph{Car} dataset all series are ascending and over the entire \emph{Music-Full} dataset 92.1\% of tuples belong to ascending series, thus, unidirectional abcODs are most common in practice.  

\begin{table}[t]\centering
\caption{\small{Statistics for experimental datasets}. 
\label{tab:stis}}
\small
\scalebox{0.7}{
\begin{tabular}{|c|c|c|c|c|c|c|c|c|}
\hline
{\sf \textbf{dataset}} & {\sf \textbf{\# tuples}} & {\sf \textbf{\# series}} & {\sf \textbf{max SS}} & {\sf \textbf{min SS}} & {\sf \textbf{avg SS}}& {\sf \textbf{\% MV}} & {\sf \textbf{\% IV}}& {\sf \textbf{\% Inc}}\\
\hline
{\em Music-Full} & 942611 & 75397 & 3052 & 1 & 12.5 & 7.8 & 1 & 92.1\\
{\em Music-Random}&1794&67&433&2&26.9&6.5&1 & 86\\
{\em Music-IncDec}&4506&43&433&6&104.8&4.7&2.6 & 80.5\\
{\em Music-Inc}&2188&25&433&6&87.5&3.5&6.0& 100\\
{\em Music-Simple}&376&1&376&376&376&7.4&3.7&100\\
{\em Car}&362& 34 &239&1&10.6&8.8&41.7&100\\
\hline

\end{tabular}
}
    \paddingD
\end{table}

\begin{table}[t]\centering
  \caption{\small{Summary of discovery methods}.
  \label{tab:impt}}
  \small
  \scalebox{0.7}{
    \begin{tabular}{| l|c|c|c|c|c|}
      \hline
      &{\sc Gap}&{\sc MonoScale}&{\sc LMS}& \sc{SD} & {\sc SD-PIE} \\
      \hline
      {\em Small Violation to Monotonicity}&&$\surd$&&$\surd$&$\surd$\\
      {\em  Outliers}&&&$\surd$&$\surd$&$\surd$\\
      \hline
    \end{tabular}
  }
    \paddingD
\end{table}

\begin{itemize}[nolistsep,leftmargin=*]
\item {\em Music-Full} is the full \emph{Music} dataset with 1M tuples.
\item {\em Music-Random} is a random sample of the above dataset by providing incomplete information from each series.
\item {\em Music-IncDec} has series with ascending \& descending orders.
\item {\em Music-Inc} contains music series with only ascending orders.
\item {\em Music-Simple} all tuples belong to a single 
  series.
\item {\em Car} contains vehicle information from multiple brands.
\end{itemize}

\smallskip\noindent\underline{CER Datasets.}  Although the
  real-world \emph{Music} dataset has real errors, we also randomly
  modify this dataset for some experiments with synthetic errors to
  control the error rate by replacing
  original values. We denote the perturbed datasets with a controlled
  error rate as CER datasets.  To evaluate the robustness, we vary the
  missing and erroneous values in the range of 5\% to 35\%.

\smallskip
\noindent{\underline{Gold Standard.}} 
We verify the ground truth as follows.
\begin{itemize}[nolistsep,leftmargin=*]
\item {\em Real-world Datasets}: 
For all real-world datasets except \emph{Music-Full}, we manually verify the correctness of series wrt (bidirectional) abcODs of all variations. For \emph{Music-Full} the values provided are estimates based on our algorithms, and annotations in the original data. This is summarized in Table~\ref{tab:stis}. 
  \item {\em CER datasets}: We use manually-verified ground truth of series with respect to (bidirectional) abcODs over real-world datasets for CER-datasets. 
\end{itemize}

\smallskip
\noindent \textbf{Algorithms.} 
We developed the following discovery algorithms in Java summarized in
Table~\ref{tab:impt}.
\begin{itemize}[nolistsep,leftmargin=*]
\item {\sc Gap}: baseline algorithm that segments data based on big gaps 
by outlier detection techniques with 3-standard deviations~\cite{BT78}.
\item {\sc MonoScale}: discovers series
  using approximate monotonicity with
  scale~\cite{DBLP:conf/ijcai/BrooksYL05}. It tolerates small
  monotonicity violations, however, does not consider outliers.
\item {\sc LMS}: discovers series by the
  concept of longest monotonic subsequences ({\sc LMS})~\cite{Golab:2009:SD:1687627.1687693}. It can
  detect erroneous values in each series, but does not allow small
  variations.
\item {\sc SD}: is our series discovery algorithm without pieces.
\item {\sc SD-PIE}: is our pieces-based solution to discover series.
\end{itemize}

Experiments were run on an Ubuntu Linux machine with an Intel(R) Xeon(R) CPU E5-2630 v3 @ 2.40GHz and 64 GB of RAM.




\subsection{Quality of abcOD Discovery} 
\label{sub:series_evaluation}

\begin{table*}[htbp]\centering
\caption{\small{Discovery quality on \emph{Music} and \emph{Car} datasets.} 
\label{table:series-all}}
\small
\scalebox{0.9}{
\begin{tabular}{|l|ccc|ccc|ccc|ccc|ccc|}
\hline
 & \multicolumn{3}{c}{\sf \textbf{GAP}} & \multicolumn{3}{|c|}{\sf \textbf{MonoScale}} & \multicolumn{3}{|c|}{\sf \textbf{A-MonoScale}} &\multicolumn{3}{|c|}{\sf \textbf{LMS}} & \multicolumn{3}{|c|}{\sf \textbf{SD-PIE}} \\
 & \textbf{F-1} & \textbf{Precision} & \textbf{Recall} & \textbf{F-1} & \textbf{Precision} & \textbf{Recall} & \textbf{F-1} & \textbf{Precision} & \textbf{Recall} & \textbf{F-1} & \textbf{Precision} & \textbf{Recall} & \textbf{F-1} & \textbf{Precision} & \textbf{Recall} \\ 
\hline
\sf\em{Music-Simple} &  
0.97&1&0.95 & 0.29&1&0.17 
&\textbf{1}&1&1 
&\textbf{1}&1&1
&\textbf{1}&1&1\\ 
\hline
\sf\em{Music-Inc} &  
0.86&0.79&0.95 
& 0.33&0.94&0.20 &0.79&0.97&0.67 
&\textbf{0.99}&0.99&0.99 
&\textbf{0.99}&0.99&1\\
\hline
\sf\em{Music-IncDec} &  0.77&0.63&0.98 & 0.46&0.83&0.32 &0.80&0.91& 0.72 &0.78&0.98&0.65 &\textbf{0.95}&0.94&0.95\\
\hline
\sf\em{Music-Random} &  0.73&0.58&0.99 & 0.59&0.81&0.47 &0.86&0.90& 0.82 &0.81&0.97&0.69 &\textbf{0.93}&0.94&0.93\\
\hline
\sf\em{Car} &  0.53&0.73&0.41 & 0.35&0.91&0.22 &&& &0.96&0.98&0.94 &\textbf{0.97}&0.98&0.97\\
\hline
\end{tabular}
}
    \paddingD
\end{table*}

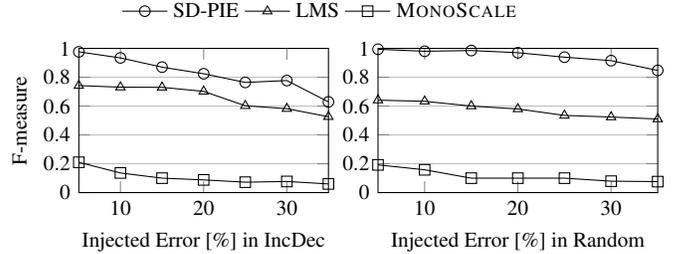
\begin{figure}[t]\centering
	\ref{legend_3}\\
	\begin{minipage}[t]{0.48\linewidth} \centering
		\begin{tikzpicture}
			\begin{axis} [
				legend to name=legend_3,
				xmax = 35,
				xmin= 5,
				ymin = 0,
				ymax = 1,
				xlabel = {Injected Error [\%] in IncDec},
        ylabel = {F-measure},
        width = 4.9 cm,
        ymajorgrids,
				]
        \pgfplotstableread{figs/music_quality_incdec.tsv}\data
			    \addplot[mark=o] table[x=error, y expr=\thisrow{f1_lmb_pie}] {\data};
			    \addlegendentry{SD-PIE}
			    \addplot[mark=triangle] table[x=error, y expr=\thisrow{f1_lms}] {\data};
			    \addlegendentry{LMS}
			    \addplot[mark=square] table[x=error, y expr=\thisrow{f1_mono}] {\data};
			    \addlegendentry{{\sc MonoScale}}
			\end{axis}
		\end{tikzpicture}
   \label{sfig:f3}
  \end{minipage}
   \hfill
 	\begin{minipage}[t]{0.48\linewidth} \centering
 		\begin{tikzpicture}
 			\begin{axis} [
 				legend to name=legend_3,
 				xmax = 35,
 				xmin= 5,
 				ymin = 0,
 				ymax = 1,
 				xlabel = {Injected Error [\%] in Random},
        ymajorgrids,
 				]
         \pgfplotstableread{figs/music_quality_random.tsv}\data
			    \addplot[mark=o] table[x=error, y expr=\thisrow{f1_lmb_pie}] {\data};
			    \addlegendentry{SD-PIE}
			    \addplot[mark=triangle] table[x=error, y expr=\thisrow{f1_lms}] {\data};
			    \addlegendentry{LMS}
			    \addplot[mark=square] table[x=error, y expr=\thisrow{f1_mono}] {\data};
			    \addlegendentry{{\sc MonoScale}}
 			\end{axis}
 		\end{tikzpicture}
    \label{sfig:f4}
   \end{minipage}
    \paddingT
   \caption{Discovery quality on {\em Music CER} datasets.\label{fig:series_mix}}
    \paddingD
  \end{figure}

\underline{\em Real-world Data.} Table~\ref{table:series-all}  presents the results of the (bidirectional) abcOD discovery on the real-world datasets. We made the following observations on the {\em Music} datasets.  {\sc Gap} achieves high recall over all datasets with a large loss in precision. As the algorithm relies on big ``gaps'' in ${\sf cat\#}$ to discover series and most catalog numbers in the same series are close enough, it only splits series occasionally. Thus, due to its simplicity, the {\sc GAP} algorithm has a high recall, however, the ``gaps'' of ${\sf cat\#}$ between consecutive series are not always large, which causes the algorithm to  merge series unnecessarily and leads to low precision. {\sc MonoScale} has a high precision and the lowest recall among all algorithms, since it does not take into account outliers and tends to split series when the outliers occur.  To overcome the flaw, we implemented a version of the algoirthm called {\sc A-MonoScale} that iteratively removes single outliers to discover series.  As shown in Table~\ref{table:series-all},  the adapted method increases recall over all datasets. {\sc LMS} is tolerant to outliers in each series, however, does not handle small violations to monotonicity (treating them as outliers), hence, it achieves high precision by splitting series (due to many consecutive outliers detected).  Finally, our
{\sc SD-PIE} approach (and thus, also our {\sc SD} approach) overcomes the problems of other techniques. It dominates other approaches 
(F-measure above 0.93 and improved by up to 17\% over other methods) and also achieves high accuracy and recall
in all datasets (significantly better than other methods). We made similar observations over the results for the {\em Car} dataset (Table~\ref{table:series-all}).
Based on Theorem~\ref{theo2:greedy} {\sc SD-PIE} achieves the same accuracy as {\sc SD}
over {\em Car} as all series in this dataset are
ascending.

\noindent
\underline{\em CER Datasets}: 
Fig.~\ref{fig:series_mix} illustrates the quality results of abcOD discovery on the
{\em Music} CER datasets. We observe analogous behaviors of all 
approaches with the controlled error rate as on the real datasets. The algorithm  achieves high F-measure  (above .82) for a reasonable amount of noise (up to 15\%). 

\subsection{Band-Width Variations}

\begin{figure}[t]\centering
	\ref{legend_6}\\
	\begin{minipage}[t]{0.48\linewidth} \centering
		\begin{tikzpicture}
			\begin{axis} [
				legend to name=legend_6,
				xmax = 20,
				xmin= 0,
				ymin = 0.2,
				ymax = 1,
				xlabel = {$\Delta$ in Music-IncDec Dataset},
        ylabel = {F-measure},
        width = 4.9 cm,
        ymajorgrids,
				]
        \pgfplotstableread{figs/varying_delta_incdec.tsv}\data
			    \addplot[mark=o] table[x=delta, y expr=\thisrow{f1_lmb_pie}] {\data};
			    \addlegendentry{SD-PIE}
			    \addplot[mark=triangle] table[x=delta, y expr=\thisrow{f1_mono}] {\data};
			    \addlegendentry{MonoScale}
			\end{axis}
		\end{tikzpicture}
   \label{sfig:f3}
  \end{minipage}
   \hfill
 	\begin{minipage}[t]{0.48\linewidth} \centering
 		\begin{tikzpicture}
 			\begin{axis} [
 				legend to name=legend_6,
 				xmax = 20,
 				xmin= 0,
 				ymin = 0.4,
 				ymax = 1,
 				xlabel = {$\Delta$ in Random Dataset},
        ymajorgrids,
 				]
         \pgfplotstableread{figs/varying_delta_random.tsv}\data
			    \addplot[mark=o] table[x=delta, y expr=\thisrow{f1_lmb_pie}] {\data};
			    \addlegendentry{SD-PIE}
			    \addplot[mark=triangle] table[x=delta, y expr=\thisrow{f1_mono}] {\data};
			    \addlegendentry{MonoScale}
 			\end{axis}
 		\end{tikzpicture}
    \label{sfig:f4}
   \end{minipage}
    \paddingT
   \caption{Discovery when varying band-width $\Delta$.\label{fig:delta_accuracy}}
    \paddingD
  \end{figure}
  
  \begin{figure}[t]\centering
	\ref{legend_5}\\
	\begin{minipage}[t]{0.48\linewidth} \centering
		\begin{tikzpicture}
			\begin{axis} [
				legend to name=legend_5,
				xmax = 100,
				xmin= 10,
				ymin = 1,
				ymode = log,
				ymax = 1000000,
				xlabel = {\% of tuples},
        ylabel = {Runtime [ms]},
        width = 4.9 cm,
        ymajorgrids,
				]
        \pgfplotstableread{figs/music_abcod-runtime.tsv}\data
			    \addplot[mark=o] table[x=ratio, y expr=\thisrow{lmb_opt}] {\data};
			    \addlegendentry{SD}
			    \addplot[mark=triangle] table[x=ratio, y expr=\thisrow{lmb_pie}] {\data};
			    \addlegendentry{SD-PIE}
			\end{axis}
		\end{tikzpicture}
   \label{sfig:f3}
  \end{minipage}
   \hfill
 	\begin{minipage}[t]{0.48\linewidth} \centering
 		\begin{tikzpicture}
 			\begin{axis} [
 				legend to name=legend_5,
				xmax = 100,
				xmin= 10,
				ymin = 0.8,
				ymax = 1,
				xlabel = {\% of tuples},
        ylabel = {F-measure},
        width = 4.9 cm,
        ymajorgrids,
 				]
         \pgfplotstableread{figs/music_abcod-fmeasure.tsv}\data
			   \addplot[mark=o] table[x=ratio, y expr=\thisrow{lmb_opt}] {\data};
			    \addlegendentry{SD}
			    \addplot[mark=triangle] table[x=ratio, y expr=\thisrow{lmb_pie}] {\data};
			    \addlegendentry{SD-PIE}
 			\end{axis}
 		\end{tikzpicture}
    \label{sfig:f4}
   \end{minipage}
    \paddingT
   \caption{SD vs. SD-PIE on Music-IncDec dataset.\label{fig:music_opt_greedy}}
   \paddingD
  \end{figure}

  Note that we manually specify band-width parameter only in this
  subsection to evaluate the effect of the parameter variations.  We
  compare the results of our (bidirectional) abcOD discovery solution on the
  real-world \emph{Music} datasets with {\sc MonoScale} for which
  band-width also plays a role. Our solution dominates {\sc
    MonoScale} in terms of F-measure as
  reported in Fig.~\ref{fig:delta_accuracy}. \eat{Additionally,
    we report for our {\sc SD-PIE} solution precision and
    recall. (Although we do not report precision and recall in the
    figure for {\sc MonoScale} the results are also dominated by {\sc
      SD-PIE}.)} The recall of the algorithm tends to decrease when
  band-width increases (not shown in
    Fig.~\ref{fig:delta_accuracy}). This is because as band-width
  increases, the method is more tolerant to violations,
  which leads to wrongly merging series.

  Our solution achieves the best F-measure when $\Delta= 3$ on the
  {\em Music} dataset. This is because $\sf year$ denotes
    release date of the records, ${\sf cat\#}$ is assigned to a record
    at early stages of the production (Section~\ref{sec:mot}), and the
    lifespan of producing music records varies from a short period of
    time to up to a few years based on the complexity of the product
    and available resources. Our algorithm for automatically
  discovering the band-width parameter described in
  Sec~\ref{sub:estimating_parameters} finds the right band-width.
%
We also observed that increasing band-width leads to a lower runtime,
because as the band-width increases, the LMBs become longer.


\subsection{Efficiency and Effectiveness} \label{exp:effect}

\begin{figure}[tbp]\centering
	\ref{legend_1}\\
		\begin{tikzpicture}
			\begin{axis} [
				legend to name=legend_1,
				legend columns= -1,
				xmax = 10,
				xmin= 1,
				ymin = 100,
				ymax = 100000000,
				ymode=log,
				xlabel = {Number of tuples * 50k},
				ylabel = {Runtime [ms]},
        ymajorgrids,
				]
        \pgfplotstableread{figs/series_scalability_500K.tsv}\data
			    \addplot[mark=triangle] table[x=ratio, y expr=\thisrow{gap}] {\data};
			    \addlegendentry{GAP}
			    \addplot[mark=*] table[x=ratio, y expr=\thisrow{lmb_pie}] {\data};
			    \addlegendentry{SD-PIE}
			    \addplot[mark=square] table[x=ratio, y expr=\thisrow{lms}] {\data};
			    \addlegendentry{LMS}
			    \addplot[mark=triangle*] table[x=ratio, y expr=\thisrow{lmb_opt}] {\data};
			    \addlegendentry{SD}
			\end{axis}
		\end{tikzpicture}
   \label{sfig:f1}
    \paddingT
   \caption{Runtime on {\em Music} Dataset. \label{fig:runtime}}
    \paddingD
  \end{figure}

  We evaluate the scalability of the different discovery algorithms
  over 500K tuples fraction of the {\em Music-Full} dataset divided
  into 10 random portions in Fig.~\ref{fig:runtime}.
  We observe that (1) the pieces-based {\sc SD-PIE} algorithm
    significantly reduces the runtime over the {\sc SD}
    algorithm on average by two orders-of-magnitude, however, without
    sacrificing the accuracy over bidirectional abcODs as illustrated in
    Fig.~\ref{fig:music_opt_greedy} over the {\em Music-IncDec}
    dataset divided into 10 portions.
   The runtime is a consequence of the complexity of the (bidirectional) abcOD discovery problem, which for {\sc SD} is $O(n^3 \log n)$ in the number of tuples (Theorems~\ref{theo:series} and~\ref{thm:optBid}). We developed pruning strategies in the {\sc
      SD-PIE} algorithm that is $O(n \log n)$ in the number of tuples ($n$) multiplied by $O(m^2)$ in the
    number of pieces ($m$). Note that
    in practice pieces are large, hence, the number of pieces is small
    (i.e., $m \ll n$); (2) {\sc SD-PIE} has smaller runtime
  than that of {\sc LMS} because it generates a smaller number of
  pieces as {\sc LMS} does not allow for small variations; (3) 
    {\sc MonoScale} (not shown in Fig.~\ref{fig:runtime}) has
    comparable runtime to {\sc SD-PIE}; while {\sc GAP} is faster
    than {\sc SD-PIE}, due to its simplicity by relying on large
    gaps, it has a much worse accuracy as reported in
    Section~\ref{sub:series_evaluation}.

\eat{
\begin{figure}[t]  \centering
  \includegraphics[scale=0.28]{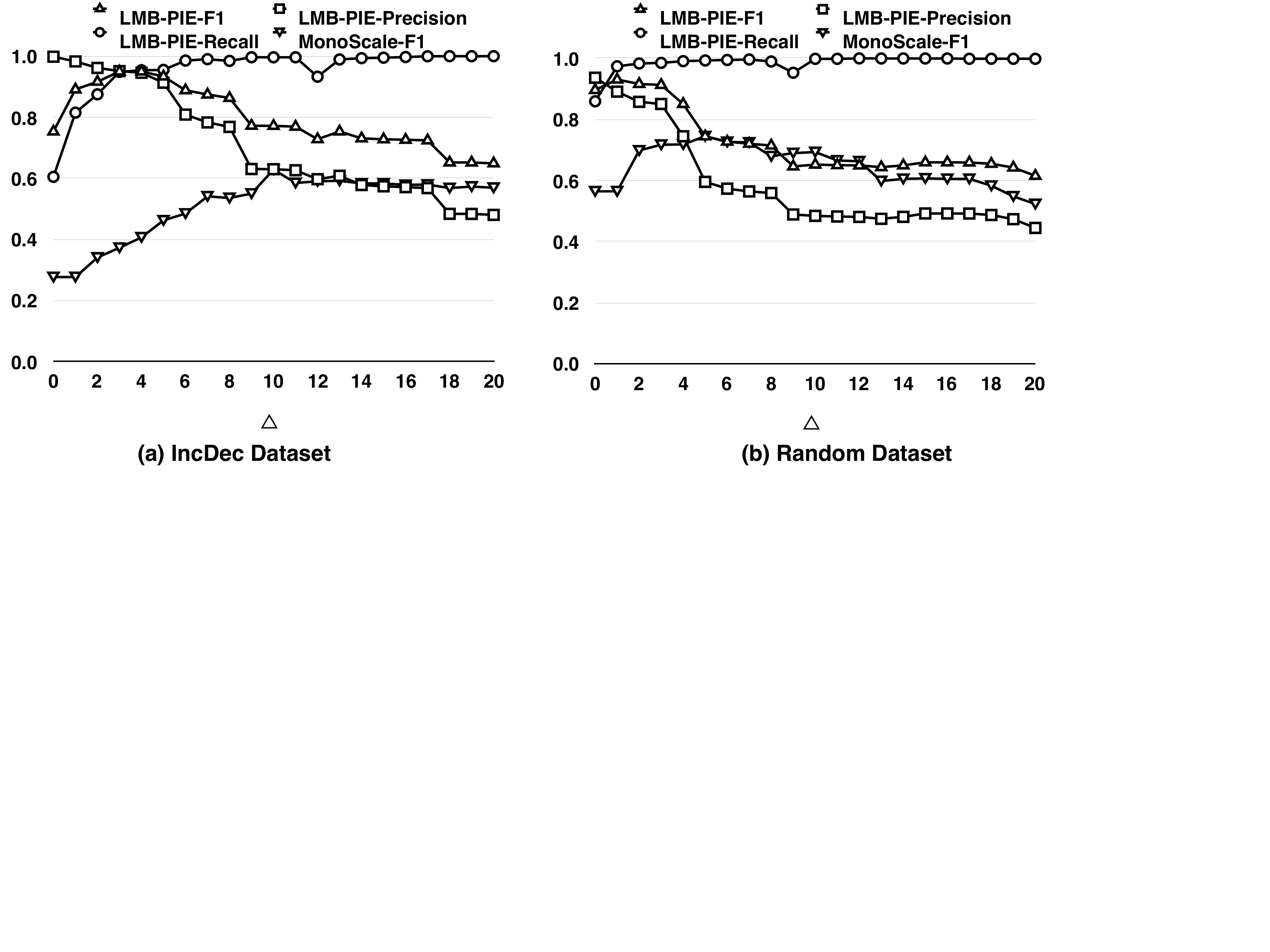}
  \caption{Discovery when varying band-width $\Delta$.\label{fig:delta_accuracy}}
\end{figure}
}

The {\sc SD-PIE} discovery algorithm runs over the {\em Music-Full}
dataset with 1M tuples for 31.5 minutes.\eat{
This is reasonable since data profiling is a periodic
  task. The data discovery engine can be run offline inside the
  organization, when the resources over the systems are not in use, or
  when the load is low.  This includes nights, and other non-peak
  hours, such as weekend and holidays.
Data dependency discovery is known to be a hard and computationally
expensive problem for other types of data dependencies including
various variations of functional dependencies~\cite{Golab:2008:GNT:1453856.1453900,FGL11}, order
dependencies~\cite{Langer:2016:EOD:2907337.2907581,DBLP:journals/pvldb/SzlichtaGGKS17,SGG18},
sequential dependencies~\cite{Golab:2009:SD:1687627.1687693} and
denial constraints~\cite{CIP13}.
}
Note again that {\sc SD-PIE} achieves the same accuracy as {\sc SD}
over {\em Car} as all series in this dataset are
increasing (Theorem~\ref{theo2:greedy}).

\subsection{Discovery over Multiple Attributes}\label{appendix:multiple}

To measure the effectiveness and efficiency of the (bidirectional) abcOD discovery
over multiple attributes, we use both the \emph{Music} and \emph{Car}
datasets to generate the following data.

\begin{itemize}[nolistsep,leftmargin=*]
\item {\em 2-Attributes}: Attribute ${\sf year}$ is split into ${\sf centuries}$ and ${\sf years}$ (e.g., $1993$ is $19$ and $93$). 
\item {\em 4-Attributes}: Attribute ${\sf year}$ is split into: ${\sf millenniums}$, ${\sf centuries}$, ${\sf decades}$ and ${\sf years}$ (e.g., $1993$ is $1$, $9$, $9$ and $3$). 
\end{itemize}

Table~\ref{table:multi-attributes} shows the results of the (bidirectional) abcOD discovery.
We observe that our solution over multiple attributes obtains similar F-measure as over a single attribute (i.e., ${\sf year}$) in both datasets. Furthermore, our solution over four attributes has slightly lower F-measure, because the distance function leads to slightly different value when ${\sf year}$ is split into four attributes.

\begin{figure}[tbp]\centering
	\ref{legend_8}\\
		\begin{tikzpicture}
			\begin{axis} [
				legend to name=legend_8,
				xmax = 4,
				xmin= 1,
				ymin = 10,
				ymax = 1000000,
				ymode=log,
				xlabel = {Number of attributes},
				ylabel = {Runtime [ms]},
        ymajorgrids,
				]
        \pgfplotstableread{figs/multi_attribute_runtime.tsv}\data
			    \addplot[mark=triangle] table[x=attribute_count, y expr=\thisrow{simple}] {\data};
			    \addlegendentry{Music-Simple}
			    \addplot[mark=*] table[x=attribute_count, y expr=\thisrow{inc}] {\data};
			    \addlegendentry{Music-Inc}
			    \addplot[mark=square] table[x=attribute_count, y expr=\thisrow{incdec}] {\data};
			    \addlegendentry{Music-IncDec}
			    \addplot[mark=triangle*] table[x=attribute_count, y expr=\thisrow{random}] {\data};
			    \addlegendentry{Music-Random}
			     \addplot[mark=diamond] table[x=attribute_count, y expr=\thisrow{car}] {\data};
			    \addlegendentry{Car}
			\end{axis}
		\end{tikzpicture}
   \label{sfig:f1}
    \paddingT
   \caption{Runtime {\em multi-attributes}.\label{fig:multi-attributes-runtime}}
    \paddingD
  \end{figure}
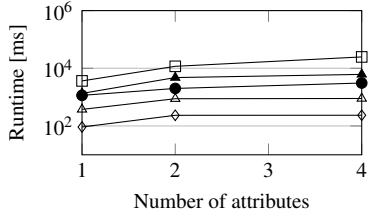

Running (bidirectional) abcOD discovery over multiple attributes takes as expected more time (however, reasonably more) than that on a single attribute (Figure~\ref{fig:multi-attributes-runtime}). We made similar observations by considering other attributes that cannot be computed from $\sf{year}$ over the \emph{Music} dataset, such as the categorical attribute $\sf{month}$, over a band OD $\sf{cat\#} \mapsto_{\Delta} \mbox{[}\sf{year}, \sf{month}\mbox{]}\upA$. The time overhead over a band OD $\sf{cat\#} \mapsto_{\Delta} \sf{year}\upA$ is around 40\% on average.

\subsection{Candidate Generation}\label{sec:generation}

We measured that the divide-and-conquer approach (described in Section~\ref{sub:estimating_parameters}) based on traditional approximate ODs to identify candidates for embedded bandODs leads to an increased number of reported ``errors'' without there often being an actual violation. The error rates for the \emph{Music} dataset are 15\% and 20\% with the (bidirectional) abcOD and traditional OD discovery, respectively (the error rates for the \emph{Car} dataset are 15\% and 23\%, respectively). We verified that embedded band ODs, ${\sf cat\#}$ $\mapsto_{\Delta}$ $\overline{{\sf year}}$ over the \emph{Music} dataset and ${\sf VIN}$ $\mapsto_{\Delta}$ $\overline{{\sf year}}$ over the \emph{Car} dataset, identified by the global traditional OD discovery by divide-and-conquer method (Sec.~\ref{sub:data_segmentation_into_series}) ranked by the measure of interestingness are indeed the most interesting.


\eat{
\begin{figure}[htbp] \centering
  \includegraphics[scale=0.26]{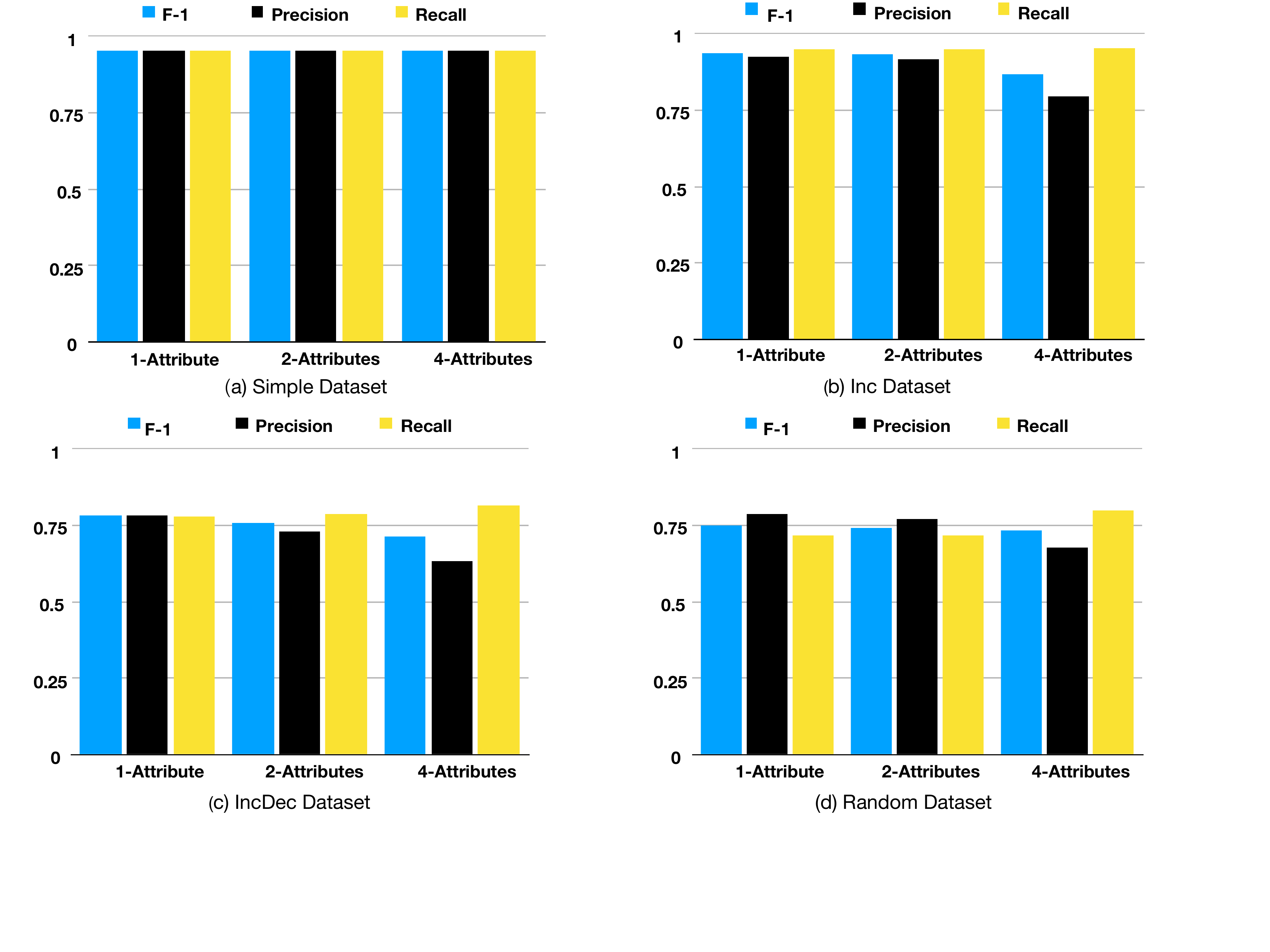}
  \caption{Discovery with multiple attrs on \emph{Music} dataset}.
  \label{fig:multi-attributes-music}
\end{figure}
}

\begin{table}[tbp]\centering
\caption{\small{Discovery multiple attributes on \emph{Music} and \emph{Car}}.
\label{table:multi-attributes}}
\small
\scalebox{0.85}{
\begin{tabular}{|l|ccc|ccc|ccc|}
\hline
 & \multicolumn{3}{c}{\sf \textbf{1-Attribute}} & \multicolumn{3}{|c|}{\sf \textbf{2-Attributes}} & \multicolumn{3}{|c|}{\sf \textbf{4-Attributes}}\\
 & \textbf{F-1} & \textbf{Pre.} & \textbf{Recall} & \textbf{F-1} & \textbf{Pre.} & \textbf{Recall} &  \textbf{F-1} & \textbf{Pre.} & \textbf{Recall} \\ 
\hline
\sf\em{Simple}
&1&1&1
&1&1&1
&1&1&1\\
\hline
\sf\em{Inc}
&\textbf{0.99}&0.99&1 
&0.98&0.96&0.99
&0.91&0.84&1\\
\hline
\sf\em{IncDec} 
&  \textbf{0.95}&0.94&0.95 
& 0.93&0.90&0.95 
&0.88&0.80& 0.98 \\
\hline
\sf\em{Random} 
&  \textbf{0.93}&0.94&0.93 
& 0.91&0.94&0.88 
&0.90&0.84& 0.97 \\
\hline
\sf\em{Car} 
&  \textbf{0.97}&0.98&0.97 
& \textbf{0.97}&0.98&0.97 
&0.96&0.98& 0.94\\
\hline
\end{tabular}
}
    \paddingD
\end{table}

%% file: related.tex
\section{Related Work} \label{sec:related_work}

Integrity constraints that specify attribute relationships, are commonly used to characterize data quality. Functional dependencies (FDs) are one of the oldest and most popular type of integrity constraints~\cite{DBLP:journals/cj/HuhtalaKPT99}. 
FDs have found application in a number of areas, such as schema design, data integration, query optimisation, data cleaning, and security~\cite{FGL11}.
In practice, dependencies may not hold exactly, due to errors in the data. Thus, approximate FDs have been defined that hold with some exceptions controlled by the number of tuples to be removed from the given table for the dependency to be satisfied. To effectively identify data quality rules different techniques to discover approximate FDs~\cite{DBLP:journals/cj/HuhtalaKPT99,PN16}, and conditional FDs that hold over subsets of the data, have been developed~\cite{Golab:2008:GNT:1453856.1453900,FGL11}.

A number of extensions to the classical notion of FDs have been proposed to express monotonicity including order dependencies (ODs)~\cite{Langer:2016:EOD:2907337.2907581,Szlichta:2012:FOD:2350229.2350241,Szlichta:2013:ECO:2556549.2556568,DBLP:journals/pvldb/SzlichtaGGKS17,SGG18} that subsume FDs. While discovery of a specified OD can be performed in linear time in the number of tuples, the discovery of a predefined approximate OD raises the complexity to quadratic in the number of attributes for both unidirectional ODs and bidirectional ODs that consider a mix of ascending and descending order~\cite{SGG18}. However, the prior work on discovery of approximate ODs~\cite{Langer:2016:EOD:2907337.2907581,DBLP:journals/pvldb/SzlichtaGGKS17,SGG18} does not consider discovering conditional dependencies that hold on subsets of the data (and small variations) as in our work, which is an involved process. 

%


Different variations of ODs have been studied including sequential dependencies (SDs)~\cite{Golab:2009:SD:1687627.1687693}.
SDs specify that when tuples have consecutive antecedent values, their consequents must be within a specified  range. The discovery problem was studied for approximate and conditional SDs~\cite{Golab:2009:SD:1687627.1687693}. However, in contrast to abcODs, SDs do not allow for small variations controlled by band-width, and a mix of ascending and descending trends.
Denial Constraints (DCs)~\cite{CIP13}, a universally quantified first-order logic formalism, are more expressive than ODs~\cite{Szlichta:2013:ECO:2556549.2556568}. DCs subsume ODs as they allow six operators for comparison of tuples $\mathcal{\{ <, >, \leq, \geq, =, \not = \} }$. The authors in ~\cite{CIP13} study the discovery of approximate DCs without considering conditional dependencies. Also, abcODs express order with small variations causing DCs to be violated without actual violation of application semantics.
Differential dependencies (DDs)~\cite{Song:2011:DDR:2000824.2000826} require that if the values of the attribute from antecedent of the DD satisfy a distance constraint, then the values of attributes from the consequent of the DD satisfy a distance constraint. The discovery problem for approximate and conditional DDs that hold approximately over subsets of the data have not been studied.

Our problem is relevant to a sequence segmentation~\cite{DBLP:conf/sdm/TerziT06} into non-overlapping partitions characterized by a model (e.g., mean and median), a general data
mining problem for
analyzing sequential data. 
Solutions to a sequence segmentation
fall into two categories~\cite{DBLP:conf/sdm/TerziT06}: (1) fast
heuristic algorithms, including top-down~\cite{Lavrenko00miningof, DBLP:conf/icai/WuCE10}, bottom-up~\cite{Palpanas:2004:OAA:977401.978140,DBLP:conf/sdm/TerziT06} and randomized~\cite{DBLP:conf/icdm/HimbergKMTT01} greedy and (2) approximation algorithms~\cite{Guha:2001:DH:380752.380841,DBLP:conf/sdm/TerziT06}. Existing sequence segmentation solutions do not consider approximate monotonic segments and allowing for small variations.
\section{Conclusions} \label{sec:conclusion} 

We devise techniques to efficiently and effectively discover a novel data quality rule in the form of (bidirectional) abcODs. 
%
%
In future work, we plan to adapt sampling techniques used for functional dependency and key discovery~\cite{PN16} and utilize distributed computing as in previous work on data discovery that includes order operators for ODs~\cite{SGI11} to further improve the efficiency of our discovery algorithm. We are also interested in studying the properties for abcODs including axiomatization and inference~\cite{DBLP:journals/pvldb/SzlichtaGGKS17,Szlichta:2012:FOD:2350229.2350241}.